\documentclass[geosciences,article,accept,moreauthors,pdftex,10pt,a4paper]{mdpi}

\DeclareUnicodeCharacter{1E45}{\.{n}}
\DeclareUnicodeCharacter{1E41}{\.{m}}
\DeclareUnicodeCharacter{2003}{\quad}
\DeclareUnicodeCharacter{0177}{\^{y}}
\DeclareUnicodeCharacter{101}{\={a}}
\DeclareUnicodeCharacter{2009}{\thinspace}
\DeclareUnicodeCharacter{2002}{\enspace{}}
\DeclareUnicodeCharacter{2005}{\thinspace}
\DeclareUnicodeCharacter{0263}{\textipa{G}}
\DeclareUnicodeCharacter{117}{\.{e}}
\DeclareUnicodeCharacter{A0}{~}
\DeclareUnicodeCharacter{2460}{\textcircled{\scriptsize{1}}}
\DeclareUnicodeCharacter{2461}{\textcircled{\scriptsize{2}}}
\DeclareUnicodeCharacter{2462}{\textcircled{\scriptsize{3}}}
\DeclareUnicodeCharacter{2463}{\textcircled{\scriptsize{4}}}
\DeclareUnicodeCharacter{2464}{\textcircled{\scriptsize{5}}}
\DeclareUnicodeCharacter{2465}{\textcircled{\scriptsize{6}}}
\DeclareUnicodeCharacter{2466}{\textcircled{\scriptsize{7}}}
\DeclareUnicodeCharacter{2467}{\textcircled{\scriptsize{8}}}
\DeclareUnicodeCharacter{2468}{\textcircled{\scriptsize{9}}}
\DeclareUnicodeCharacter{02C2}{<}
\DeclareUnicodeCharacter{2033}{''}
\DeclareUnicodeCharacter{0229}{\c{e}}
\DeclareUnicodeCharacter{0218}{\cb{S}}
\DeclareUnicodeCharacter{0219}{\cb{s}}
\DeclareUnicodeCharacter{021B}{\cb{t}}
\firstpage{1}

\articlenumber{325}

\pubvolume{8}
\issuenum{9}
\pubyear{2018}
\copyrightyear{2018}

\history{Received: 30 July 2018; Accepted: 25 August 2018; Published: 29 August 2018}

\updates{yes} 
\pdfoutput=1

\usepackage{xcolor, colortbl}
\usepackage[normalem]{ulem}
\usepackage{longtable}
\usepackage{threeparttable}
\usepackage{soul,booktabs,multirow}
\usepackage{attrib}
\usepackage{epstopdf}
\usepackage{subfigure}
\usepackage{chemscheme}
\usepackage{mathcomp}
\usepackage{amsfonts}
\usepackage{amssymb}
\usepackage{CJKutf8}

\usepackage[T5,T1]{fontenc}

\let\originalleft\left
\let\originalright\right
\renewcommand{\left}{\mathopen{}\mathclose\bgroup\originalleft}
\renewcommand{\right}{\aftergroup\egroup\originalright}

\let\orignewcommand\newcommand  
\let\newcommand\providecommand  
\RequirePackage{verse}
\let\newcommand\orignewcommand  




\newsavebox\foobox

\renewcommand{\dagger}{\mathchar"2279}
\renewcommand{\ddagger}{\mathchar"227A}

\graphicspath{{./figures/}}
\usepackage{enumitem}
\usepackage{tabulary}
\usepackage{upgreek}
\usepackage{wasysym}
\setitemize{parsep=6pt,itemsep=0pt,leftmargin=*,labelsep=5.5mm,align=parleft}
\setenumerate{parsep=6pt,itemsep=0pt,leftmargin=*,labelsep=5.5mm,align=parleft}
\setlist[description]{itemsep=0mm}

\usepackage{fancyvrb}
\usepackage[scaled=0.85]{beramono}

\usepackage{seqsplit}
\usepackage{bm}
\usepackage{pifont}

  \newlength{\cellWidtha}
  \newlength{\cellWidthb}
  \newlength{\cellWidthc}
  \newlength{\cellWidthd}
  \newlength{\cellWidthe}
  \newlength{\cellWidthf}
  \newlength{\cellWidthg}

\newcommand{\fig}[1]{Figure~\ref{#1}}
\newcommand{\sect}[1]{Section~\ref{#1}}
\newcommand{\tabref}[1]{Table~\ref{#1}}

\newcommand{\app}[1]{Appendix~\ref{#1}}

\usepackage{cleveref}
\crefname{figure}{Figure}{Figures}
\crefname{table}{Table}{Tables}
\crefrangelabelformat{figure}{#3#1#4--#5#2#6}
\crefrangelabelformat{table}{#3#1#4--#5#2#6}
\crefname{section}{Section}{Sections}
\crefname{paragraph}{Section}{Sections}
\crefrangelabelformat{section}{#3#1#4--#5#2#6}
\crefname{appendix}{Appendix}{Appendixs}
\crefrangelabelformat{appendix}{#3#1#4--#5#2#6}
\crefname{scheme}{Scheme}{Schemes}
\crefrangelabelformat{scheme}{#3#1#4--#5\crefstripprefix{#1}{#2}#6}
\crefname{chart}{Chart}{Charts}
\crefrangelabelformat{chart}{#3#1#4--#5\crefstripprefix{#1}{#2}#6}

\newcommand{\PreserveBackslash}[1]{\let\temp=\\#1\let\\=\temp}

\usepackage{tikz}

\usepackage{setspace}
\usepackage[russian,english]{babel}
\usepackage{microtype}
\usepackage{combelow}

\makeatletter
\g@addto@macro{\UrlBreaks}{\UrlOrds}
\makeatother

\robustify{\footnote}

\Title{A Quantitative Comparison of Exoplanet~Catalogs}

\Author{Dolev~Bashi~\textsuperscript{1}\textsuperscript{,}*, Ravit~Helled~\textsuperscript{2}\href{https://orcid.org/0000-0001-5555-2652}{\orcidicon} and Shay~Zucker~\textsuperscript{1}}
\AuthorNames{Dolev Bashi, Ravit Helled and Shay Zucker}
\address{\textsuperscript{1} \quad School of Geosciences, Raymond and Beverly Sackler Faculty of Exact Sciences, Tel Aviv University, \linebreak 6997801 Tel Aviv, Israel; shayz@post.tau.ac.il

\textsuperscript{2} \quad Institute for Computational Science, Center for Theoretical Astrophysics and Cosmology, University of Zurich, Winterthurerstrasse 190, CH-8057 Zurich, Switzerland; rhelled@physik.uzh.ch}

\corres{Correspondence: dolevbas@mail.tau.ac.il; Tel.:~+972-54-7365-965}

\abstract{In this study, we investigated the differences between four commonly-used exoplanet catalogs (exoplanet.eu; exoplanetarchive.ipac.caltech.edu; openexoplanetcatalogue.com; exoplanets.org) using a Kolmogorov--Smirnov (KS) test. We found a relatively good agreement in terms of the planetary parameters (mass, radius, period) and stellar properties (mass, temperature, metallicity), although a more careful analysis of the \emph{overlap} and \emph{unique} parts of each catalog revealed some differences. We quantified the statistical impact of these differences and their potential cause. We concluded that although statistical studies are unlikely to be significantly affected by the choice of catalog, it would be desirable to have one consistent catalog accepted by the general exoplanet community as a base for exoplanet statistics and comparison with theoretical~predictions.}

\keyword{methods: statistical; astronomical data bases: miscellaneous; catalogs; planetary systems; stars:~statistics}

\makeatletter
\DeclareRobustCommand*\textsubscript[1]{%
  \@textsubscript{\selectfont#1}}
\def\@textsubscript#1{%
  {\m@th\ensuremath{_{\mbox{\fontsize\sf@size\z@#1}}}}}
\makeatother

\begin{document}
\section{Introduction \label{sect:sec1-geosciences-343110}}

Since the detection of ‘\emph{51 Peg b}’, the first exoplanet around a main sequence star~\cite{B1-geosciences-343110}, many more planets around other stars have been discovered. Currently, more than 3500 exoplanets have been detected in our galaxy. The diversity of these exoplanets in terms of orbital and physical properties is overwhelming. This diversity challenges planet formation and evolution theories, which were tuned originally to explain the planets in our Solar System~\cite{B2-geosciences-343110,B3-geosciences-343110}.

Several groups took it upon themselves to label and classify the known exoplanets, and compile catalogs to provide the scientific community with a comprehensive working tool to access the data and perform statistical studies of the exoplanet sample (hereafter, exostatistics). These databases include information about the physical properties of the planets, as well as their host stars. Analysis of this information is constantly improving our understanding of planet formation mechanisms~\cite{B4-geosciences-343110}, protoplanetary disks~\cite{B5-geosciences-343110}, and planetary composition and internal structure~\cite{B6-geosciences-343110,B7-geosciences-343110}. At the moment, several exoplanet catalogs are available and are used by the community (see~\cite{B8-geosciences-343110}).

The most widely-used exoplanet catalogs are: 
\begin{enumerate}[label=\arabic*.]
\item The Extrasolar Planets Encyclopaedia, \url{www.exoplanet.eu} (\cite{B9-geosciences-343110}; hereafter, EU).
\item \textls[-20]{The NASA Exoplanet Archive, \url{https://exoplanetarchive.ipac.caltech.edu} (\cite{B10-geosciences-343110}; hereafter, ARCHIVE).}
\item The Open Exoplanet Catalogue, \url{www.openexoplanetcatalogue.com/} (hereafter, OPEN).
\item The Exoplanet Data Explorer, \url{www.exoplanets.org} (\cite{B11-geosciences-343110}; hereafter, ORG).
\end{enumerate}

These catalogs include data from ground-based observations as well as space missions such as CoRoT, Kepler, and K2. The available data in these catalogs are comprehensive and include the physical properties of the host star, available information on the planetary physical properties, and the referenced confirmation paper or other mentioned~source.

The different teams of each catalog use different criteria to include a planet, which are usually based on the physical properties of the planet or statistical thresholds (see \tabref{tabref:geosciences-343110-t001}). Furthermore, each catalog has a different approach to displaying the database. For example, ARCHIVE designates a set of default parameters for each planet. This set is extracted from a single published reference to ensure internal consistency. Additional values published in other papers can only be found by viewing the pages dedicated to individual planets, where multiple sets of parameters are displayed. As a result, the ARCHIVE table provides a self-consistent set of parameters for any system, with missing values when the information is unavailable. On the other hand, EU uses a table displaying information on specific planet extracted from different sources, thus making for a more complete parameter set, though not necessarily~self-consistent.

Many exostatistics papers use one of these catalogs as their source of observational data. Nevertheless, so far, the different catalogs have not been compared in terms of their possible differences and potential biases and selection effects that might affect inferred results and~conclusions.
    \begin{table}[H]
    \tablesize{\small}
    \centering
    \caption{The exoplanets catalogs inclusion~criteria.}
    \label{tabref:geosciences-343110-t001}

\setlength{\cellWidtha}{\textwidth/3-2\tabcolsep-1.2in}
\setlength{\cellWidthb}{\textwidth/3-2\tabcolsep-0.4in}
\setlength{\cellWidthc}{\textwidth/3-2\tabcolsep+1.4in}
\scalebox{1}[1]{\begin{tabular}{>{\PreserveBackslash\centering}m{\cellWidtha}>{\PreserveBackslash\centering}m{\cellWidthb}>{\PreserveBackslash\centering}m{\cellWidthc}}
\toprule

\textbf{Catalog} & \textbf{Object Mass Criterion} & \textbf{Reference Criteria}\\
\cmidrule{1-3}

EU & \textless{}$60M_{J} \pm 1\sigma $  & Published or submitted to peer-reviewed journals or announced in conferences by professional astronomers.\\

ARCHIVE & \textless{}$30M_{J} $  & Accepted, refereed paper.\\

OPEN & Not listed & Open-source.\\

ORG & \textless{}$24M_{J} $  & Carefully vetted, peer-reviewed journal papers.\\

\bottomrule
\end{tabular}}

    \end{table}

In this work, we present a simple statistical comparison between the different exoplanet catalogs. We mainly focus on the EU, ARCHIVE and OPEN catalogs. The database of the ORG catalog contains a single and reliable set of parameters for each planet. However, since it has not been updated for a couple of years now (see website and discussion in Reference~\cite{B8-geosciences-343110}), we perform only a coarse comparison. As discussed in Reference~\cite{B8-geosciences-343110}, there are plans to restart regular updates in the near~future.

\section{Methods \label{sect:sec2-geosciences-343110}}

We have downloaded lists of confirmed planets from the following four catalogs: EU, ARCHIVE, OPEN and ORG on 3 April 2018. As discussed previously, because of the different planetary mass criteria of each catalog (see \tabref{tabref:geosciences-343110-t001}), we set $10M_{J} $ as an upper bound for the planetary mass, to strictly exclude any potential brown dwarfs. Thus, we avoided any biases that might emerge from the different mass cutoffs the catalogs~use.

The parameters we use in order to compare the catalogs are the stellar mass ($M_{*} $), surface temperature ($T_{eff} $) and metallicity ($\left\lbrack {Fe/H} \right\rbrack $), and planetary mass ($M_{p} $), radius ($R_{p} $) and orbital period (\emph{Period}). We chose this set of six parameters because they are the fundamental parameters that are most easily available from current photometric and spectroscopic detection methods~\cite{B12-geosciences-343110}. Physically, these parameters provide basic, broad information about the planetary system~\cite{B6-geosciences-343110}. The process of deriving the stellar properties involves a collection of literature values for atmospheric properties (temperature, surface gravity, and metallicity) derived from different observational techniques (photometry, spectroscopy, asteroseismology, and exoplanet transits), and then fitting them to stellar isochrones (e.g., References~\cite{B13-geosciences-343110,B14-geosciences-343110}). The stellar properties are then used in the derivation of almost all planet properties from radial velocities (RV), transits or transit timing variation (TTV) data. Thus, \mbox{a reliable} estimate of these parameters is crucial for the quality of the planet properties estimate (\mbox{e.g., Reference~\cite{B15-geosciences-343110}}).

In the framework of this analysis, we compared separately (and in combination) the planetary properties of the confirmed planets from the listed catalogs. In addition, we performed a comparison between planetary systems by examining the distributions of stellar and planetary properties of the main star and each system’s first detected planet. By doing so, we were able to find the biases of the planet properties emerging from the stellar properties. There was no sense in comparing the stellar parameters of all the confirmed planets since it is possible to unintentionally give more weight to multi-planetary systems when performing the analysis. \tabref{tabref:geosciences-343110-t002} lists the total number of confirmed planets and systems of each catalog as a function of the different stellar and planet properties. \mbox{The significant} variability of those numbers raised the following questions: Does the catalog with the largest number of planets include all the listed objects of the other catalogs? How different is the distribution of planets from two different~catalogs?
    \begin{table}[H]
    \tablesize{\small}
    \centering
    \caption{The total number of listed planets with different stellar and planet properties for each exoplanet~catalog.}
    \label{tabref:geosciences-343110-t002}

\setlength{\cellWidtha}{\textwidth/5-2\tabcolsep+1.6in}
\setlength{\cellWidthb}{\textwidth/5-2\tabcolsep-0.6in}
\setlength{\cellWidthc}{\textwidth/5-2\tabcolsep-0.4in}
\setlength{\cellWidthd}{\textwidth/5-2\tabcolsep-0.7in}
\setlength{\cellWidthe}{\textwidth/5-2\tabcolsep-0.6in}
\scalebox{1}[1]{\begin{tabular}{>{\PreserveBackslash\raggedright}m{\cellWidtha}>{\PreserveBackslash\centering}m{\cellWidthb}>{\PreserveBackslash\centering}m{\cellWidthc}>{\PreserveBackslash\centering}m{\cellWidthd}>{\PreserveBackslash\centering}m{\cellWidthe}}
\toprule

\multicolumn{1}{>{\PreserveBackslash\centering}m{\cellWidtha}}{\textbf{Number of Objects}} & \textbf{EU} & \textbf{ARCHIVE} & \textbf{OPEN} & \textbf{ORG}\\
\cmidrule{1-5}

All planets & 3757 & 3708 & 3524 & 2950\\

With planetary mass & 1327 & 1276 & 1275 & 2917\\

With planetary radius & 2807 & 2912 & 2731 & 2438\\

With planetary orbital period & 3505 & 3564 & 3371 & 2920\\

With planetary mass, radius and orbital period & 655 & 576 & 558 & 2436\\

All systems & 2652 & 2693 & 2556 & 2200\\

With stellar mass $M_{*} $  & 2465 & 2571 & 2476 & 1929\\

With stellar metallicity $\left\lbrack {Fe/H} \right\rbrack $  & 2444 & 2581 & 2088 & 1877\\

With stellar surface temperature $T_{eff} $  & 2503 & 2424 & 2469 & 2162\\

\bottomrule
\end{tabular}}

    \end{table}

Most of the statistical work in this analysis is based on comparing the different sets using a two-sample Kolmogorov--Smirnov test (hereafter, KS test,~\cite{B16-geosciences-343110}). Broadly speaking, the \emph{p}-value of the KS test indicates to what extent two samples can be considered to be drawn from the same distribution---a high \emph{p}-value indicates a good agreement. It is sensitive to differences in both shape and location of the empirical distribution functions of the two samples. A KS test comparison between two catalogs would compare the distributions of all available estimates of one of the planetary properties mentioned above. If a specific object’s quantity is unavailable, we excluded the object from the comparison pertaining to this property. In cases where only lower/upper bounds were available, we set it to the listed value instead of excluding it. Thus, there were some cases in which two catalogs agreed on the planetary nature of a specific object, yet, since one of the catalogs had a missing value for some property, we excluded this planet from the~test.

\begin{figure}[H]
\centering
\includegraphics[scale=1]{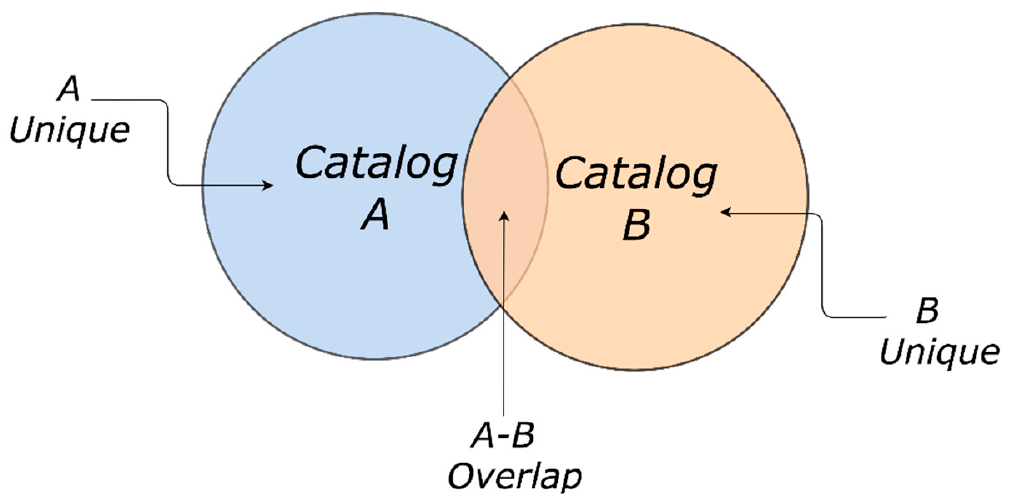}
\caption{Venn diagram describing the different compared subsets. Catalogs ‘\emph{A}’, ‘\emph{B}’ represents two out of the three compared catalogs. For each planet or stellar property we used a Kolmogorov--Smirnov (KS) test to compare two out of the three derived subsets: ‘\emph{Overlap’, ‘unique A}’ and ‘\emph{unique B}’.}
\label{fig:geosciences-343110-f001}
\end{figure}

For each pair of catalogs, ‘\emph{A}’ and ‘\emph{B}’\emph{,} we compiled three subsets: (1) The \emph{overlap} subset, including all objects listed in both catalogs; (2) the \emph{unique} ‘\emph{A}’ subset, including only the planets that are unique to catalog ‘\emph{A}’ and not listed in catalog ‘\emph{B}’; and (3) the \emph{unique ‘B}’ subset, including only the planets that are unique to catalog ‘\emph{B}’ and not listed in catalog ‘\emph{A}’. We then applied the KS test to compare the three subsets. We performed this analysis for each one of the six parameters separately, as well as a comparison of subsets that include information about the planetary mass, radius and orbital period. \fig{fig:geosciences-343110-f001} describes the methodology applied to compare the different~catalogs.

A problem we often encountered in our analysis is the possibility of different planet names and aliases listed in each of the catalogs. The differences are typically caused by spaces, uppercase/lowercase mix-up, and numbers or unique symbols that are used in spelling the stellar and planet names. Yet, in most cases, we found a given object to have two different names in two different catalogs. To overcome this difficulty, we first cross-matched the different objects according to their stellar names using the SIMBAD database, and then identified the different planetary names that populate each system. This practice imparted us with the following unfortunate conclusion: Currently, there is no consensus on an unambiguous method to mark a specific planet and to find its aliases in a convenient way. Therefore, there is no reliable way to cross-match two planetary~tables.

\section{Results \label{sect:sec3-geosciences-343110}}

\begin{figure}[H]
\centering
\includegraphics[scale=0.9]{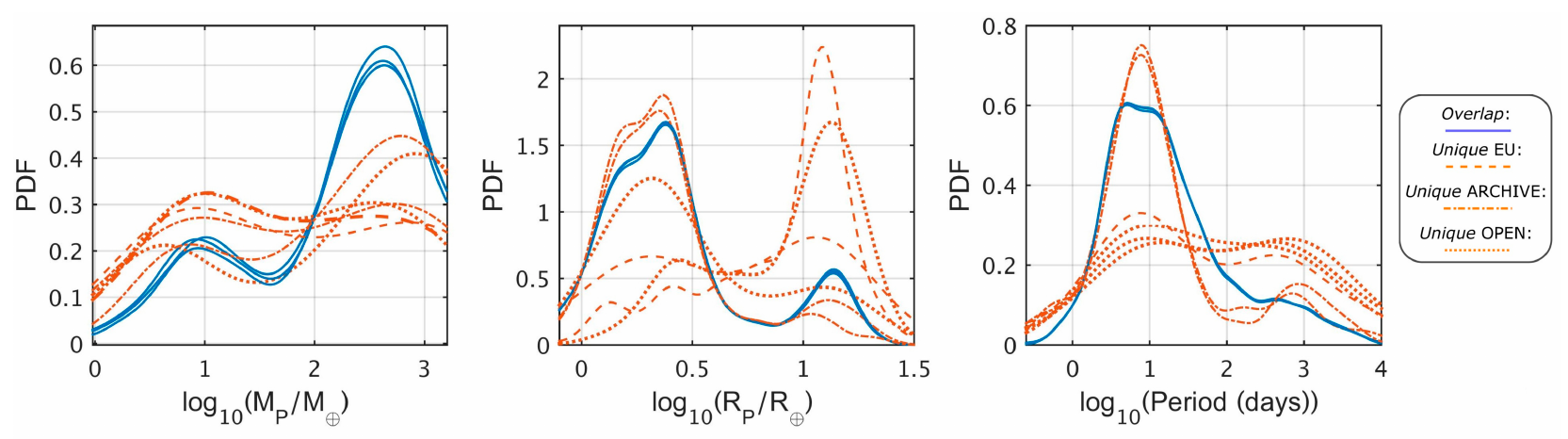}
\caption{Each catalogs\emph{}’ subsets probability density functions (PDFs) of planetary mass, radius and orbital period properties, using the kernel density estimation (KDE), where \emph{overlap} is represented by blue curves and \emph{unique} by brown curves. In each plot three \emph{overlap} subsets and six \emph{unique} subsets may be noted (see text for further details).}
\label{fig:geosciences-343110-f002}
\end{figure}

In general, we found the distributions of planets and systems from the different catalogs to be quite similar. An exception was the ORG catalog, where its missing planetary mass values were derived from some theoretical mass--radius (M--R) relation. See \app{app:app1-geosciences-343110} for the detailed and complete analysis with the quantitative results (\cref{tabref:geosciences-343110-t0A1,tabref:geosciences-343110-t0A8}; \cref{fig:geosciences-343110-f0A1,fig:geosciences-343110-f0A7}). To infer the differences between the catalogs, we compared the derived populations of the \emph{overlap} and unique subsets for two catalogs at a time, as discussed above. We found the distributions of the \emph{overlapping} planets between the different catalogs to be similar. However, there were significant differences between the \emph{overlap} and unique subsets. In \fig{fig:geosciences-343110-f002}, a kernel density estimation (KDE)~\cite{B17-geosciences-343110} for the probability density functions (PDF) of the mass, radius and orbital period for the different subsets is given. The solid blue curves represent the three \emph{overlap} subsets when comparing each pair of the EU-ARCHIVE-OPEN catalogs. The brown curves represent the unique subsets, where the dashed, dashed-dotted and dotted curves correspond to the unique EU, ARCHIVE and OPEN subsets, respectively. Evidently, the \emph{overlap} and unique subsets seem to have very different distributions for the three planet properties. \tabref{tabref:geosciences-343110-t003} presents the KS tests p-values for the different comparisons between KS tests. We found a low p-value for most comparison~tests.

As for the property of planetary mass, we noted that a higher density of small mass planets ($\left. M_{p} \right.\sim 10M_{\oplus} $) populate the unique subsets. We found most of these planets to be TTV planets, detected in multi-planetary systems. Usually, the mass of a planet detected through TTV is not resolved directly, and in practice, is degenerate with the planet\emph{}’s eccentric orbit~\cite{B18-geosciences-343110}. As a result, we have many small TTV planets with an upper mass limit, rather than a nominal value. However, there is no uniform approach to displaying this mass limit. At times, the catalogs choose to omit the value altogether, while at other times it is displayed as an upper limit. However, in many cases, especially with the EU catalog, the mass limit is reported as a valid nominal estimate. Due to this difference in the mass property criteria, we found many small mass planets biasing the unique subset towards smaller~masses.

For the planet radii and orbital period distributions, we found the EU and OPEN unique subsets to have a relative higher density of planets in radii values of $\left. R_{p} \right.\sim 1R_{J} $, and periods of \mbox{$\left. Period \right.\sim 1000\ \text{days} $}. Examining the planets that comprise these subsets suggests the reason for this difference is probably related to the different inclusion criteria the catalogs use. Some examples are: Listed planets where the confirmation paper has used some theoretical M--R relations to infer the planetary radius (or mass), some unusual large radii planets suffering strong tidal forces due to their proximity to the parent stars (in short periods), planets with ‘\emph{}strange\emph{}’ transiting light curves that make the planetary detection more controversial, etc. As for the higher consistency, in terms of the \emph{overlap} and unique parts of the ARCHIVE catalog, we found it to be somewhat artificial. By examining the ARCHIVE unique planets, we found over 75\% of them to be K2 planets, suggesting the reason for the good agreement with the \emph{overlap} subsets related to the simple fact: Most of the \emph{overlap} planets are transiting planets (Kepler and K2), with similar properties as the K2~planets.
    
    \begin{table}[H]
    \tablesize{\small}
    \centering
    \caption{The \emph{p}-values from KS tests wherein the distributions of the \emph{overlap} vs. \emph{unique} subsets of planetary mass, radius and orbital period properties are compared. The three numbers in brackets in each cell represent the \emph{p}-value according to this order: ($M_{P},~R_{P},~Period $).}
    \label{tabref:geosciences-343110-t003}

\setlength{\cellWidtha}{\textwidth/4-2\tabcolsep-0.2in}
\setlength{\cellWidthb}{\textwidth/4-2\tabcolsep+0.2in}
\setlength{\cellWidthc}{\textwidth/4-2\tabcolsep-0in}
\setlength{\cellWidthd}{\textwidth/4-2\tabcolsep-0in}
\scalebox{1}[1]{\begin{tabular}{>{\PreserveBackslash\centering}m{\cellWidtha}>{\PreserveBackslash\centering}m{\cellWidthb}>{\PreserveBackslash\centering}m{\cellWidthc}>{\PreserveBackslash\centering}m{\cellWidthd}}
\toprule

 & \textbf{\emph{Overlap} OPEN-ARCHIVE} & \textbf{\emph{Overlap} EU-OPEN } & \textbf{\emph{Overlap} ARCHIVE-EU}\\
\cmidrule{1-4}

\textbf{\boldmath{\emph{Unique} EU}} & --- & (\textless{}10\textsuperscript{$-$10}, \textless{}10\textsuperscript{$-$10}, \textless{}10\textsuperscript{$-$10}) & (\textless{}10\textsuperscript{$-$10}, \textless{}10\textsuperscript{$-$10}, \textless{}10\textsuperscript{$-$10})\\

\textbf{\boldmath{\emph{Unique} ARCHIVE}} & (3 $\times$ 10\textsuperscript{$-$8}, 0.03, 2 $\times$ 10\textsuperscript{$-$3}) & --- & (0.69, 0.01, 0.02)\\

\textbf{\boldmath{\emph{Unique} OPEN}} & (\textless{}10\textsuperscript{$-$10}, \textless{}10\textsuperscript{$-$$-$10}, 10\textsuperscript{$-$9}) & (0.034, 0.74, 4 $\times$ 10\textsuperscript{$-$8}) & ---\\

\bottomrule
\end{tabular}}

    \end{table}

\begin{figure}[H]
\centering
\includegraphics[scale=0.9]{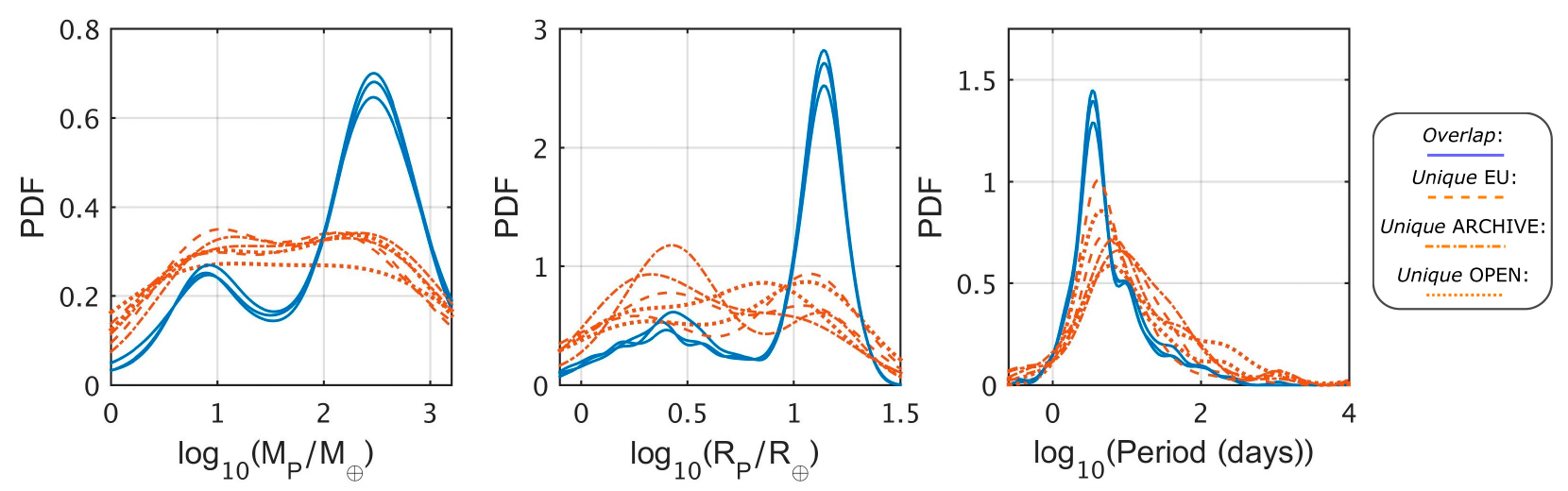}
\caption{Same as \fig{fig:geosciences-343110-f002}, but for the subset of planets in which all three properties of planetary mass, radius and orbital period were~available.}
\label{fig:geosciences-343110-f003}
\end{figure}

Performing a similar analysis for the subset of planets in which all three properties of planetary mass, radius and orbital period were available, we found the unique and overlap subsets were different again (\fig{fig:geosciences-343110-f003}), although having slightly higher p-values (\tabref{tabref:geosciences-343110-t004}) than the individual property comparisons presented~above.

In this case, most of the \emph{overlap} planets had large masses and radii, with short periods. This in itself is a bias, caused by the combined sensitivity of the RV and transit detection methods~\cite{B19-geosciences-343110}. \mbox{As for} the distributions of the \emph{unique} subsets, we found them to be compromised by a higher number of TTV-detected planets, causing the distributions to be almost uniform in the regime of both small and large planets. Another interesting aspect we observed about these \emph{unique} subsets was their relative similarity between each other, as displayed quantitative in \tabref{tabref:geosciences-343110-t0A10} displaying the unique subsets \emph{p}-values. In spite of this relative similarity, these subsets populate different planets, causing this resemblance to be a product of the TTV biases in the catalogs planet mass inclusion~criteria.
    
    \begin{table}[H]
    \tablesize{\small}
    \centering
    \caption{Same as \tabref{tabref:geosciences-343110-t003}, but for the subset of planets in which all three properties of planetary mass, radius and orbital period are~available.}
    \label{tabref:geosciences-343110-t004}

\setlength{\cellWidtha}{\textwidth/4-2\tabcolsep-0.2in}
\setlength{\cellWidthb}{\textwidth/4-2\tabcolsep+0.3in}
\setlength{\cellWidthc}{\textwidth/4-2\tabcolsep+0.2in}
\setlength{\cellWidthd}{\textwidth/4-2\tabcolsep+0.1in}
\scalebox{0.9}[0.9]{\begin{tabular}{>{\PreserveBackslash\centering}m{\cellWidtha}>{\PreserveBackslash\centering}m{\cellWidthb}>{\PreserveBackslash\centering}m{\cellWidthc}>{\PreserveBackslash\centering}m{\cellWidthd}}
\toprule

 & \textbf{\emph{Overlap} OPEN-ARCHIVE} & \textbf{\emph{Overlap} EU-OPEN } & \textbf{\emph{Overlap} ARCHIVE-EU}\\
\cmidrule{1-4}

\textbf{\boldmath{\emph{Unique} EU}} & --- & (2 $\times$ 10\textsuperscript{$-$8}, \textless{}10\textsuperscript{$-$10}, 8 $\times$ 10\textsuperscript{$-$6}) & (3 $\times$ 10\textsuperscript{$-$7}, 8 $\times$ 10\textsuperscript{$-$7}, 0.23)\\

\textbf{\boldmath{\emph{Unique} ARCHIVE}} & (3 $\times$ 10\textsuperscript{$-$5}, \textless{}10\textsuperscript{$-$10}, 3 $\times$ 10\textsuperscript{$-$11}) & --- & (0.02, 3 $\times$ 10\textsuperscript{$-$7}, 8 $\times$ 10\textsuperscript{$-$4})\\

\textbf{\boldmath{\emph{Unique} OPEN}} & (6 $\times$ 10\textsuperscript{$-$5}, 7 $\times$ 10\textsuperscript{$-$5}, 6 $\times$ 10\textsuperscript{$-$4}) & (6 $\times$ 10\textsuperscript{$-$3}, 2 $\times$ 10\textsuperscript{$-$7}, 10\textsuperscript{$-$3}) & ---\\

\bottomrule
\end{tabular}}

    \end{table}

\begin{figure}[H]
\centering
\includegraphics[scale=0.9]{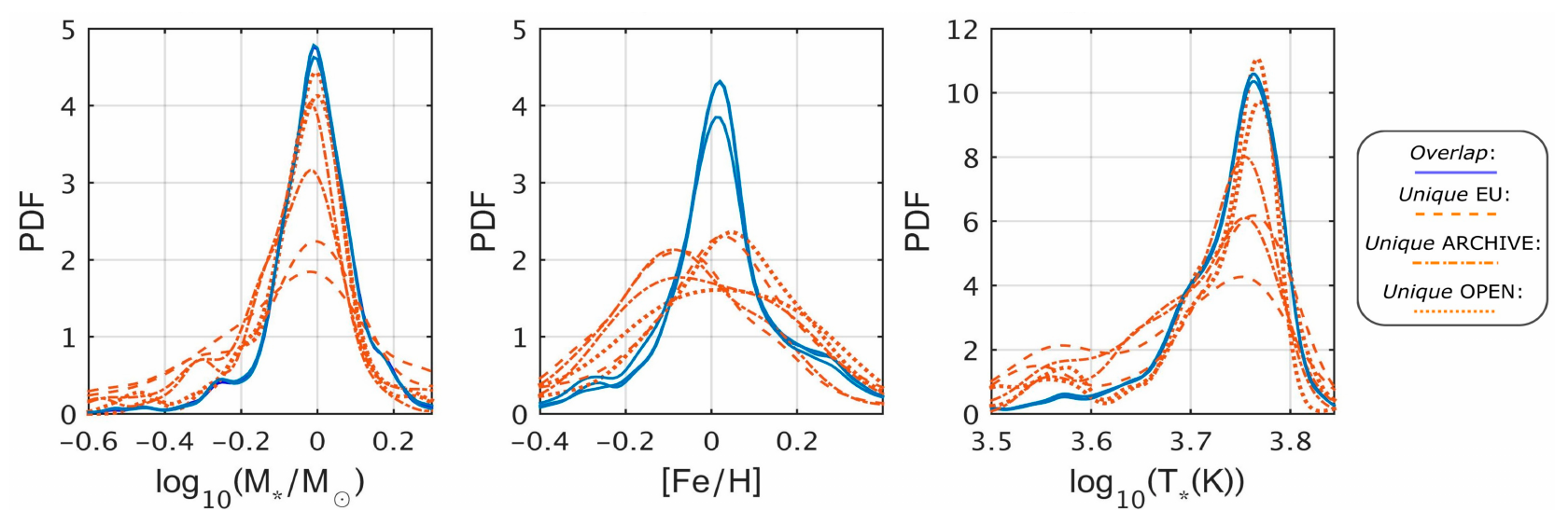}
\caption{Catalogs’ subsets PDF’s of stellar mass, metallicity and surface temperature properties using KDE, where: \emph{overlap} is given in blue and \emph{unique} is given in brown. In each plot, three \emph{overlap} subsets and six \emph{unique} subsets may be noted (see text for further details).}
\label{fig:geosciences-343110-f004}
\end{figure}

After examining the planetary systems according to the stellar properties, we again found similarity between the \emph{overlap} subsets and significant differences between the \emph{unique} to \emph{overlap} subsets (\fig{fig:geosciences-343110-f004}). We noted that the \emph{unique} distributions of stellar mass and surface temperature were biased towards smaller mass and lower temperature stars (K, M-stars). Spectroscopy of small stars (especially M-stars) is challenging because of their intrinsic faintness and high activity~\cite{B20-geosciences-343110}. As a result, detection of planets around these stars is purportedly more difficult and somewhat controversial, thus explaining why we found a higher relative number of these objects in the \emph{unique} subsets. \mbox{As for} the stellar metallicity, we found it to be an unreliable property when analyzing possible biases in the catalog comparison. Although the metallicity is an important property, providing an early record of the chemical composition of the initial protoplanetary disk~\cite{B21-geosciences-343110}, the planetary detection methods do not rely on this property directly. The catalogs usually reported this property with high relative errors, probably linked to the imprecise derivation that is used to determine the metallicity value. Adding to this the fact that each catalog referred to different stellar survey sources, we found the highest inconsistency between catalogs to be in this parameter (when compared with the other stellar and planetary properties). Nevertheless, we still found the following trends: The \emph{unique} subsets that included larger planets, especially in terms of the OPEN catalog, was biased towards higher metallicity values, as expected from previous studies~\cite{B22-geosciences-343110,B23-geosciences-343110}. On the other hand, small planets, especially listed in the unique ARCHIVE catalog, had a wider distribution of metallicities~\cite{B24-geosciences-343110}. As before, the \emph{p}-values for the different KS tests of stellar properties in the \emph{overlap} vs. \emph{unique} subsets are provided in \tabref{tabref:geosciences-343110-t005}. The seemingly improved mean \emph{p}-value results of the OPEN catalog are caused by its relative smaller number of listed stellar property~objects.
\unskip
    
    \begin{table}[H]
    \tablesize{\small}
    \centering
    \caption{The \emph{p}-values obtained from KS tests comparing the distributions of the \emph{overlap} vs. \emph{unique} subsets of stellar mass, metallicity and surface temperature properties. The three numbers in brackets of each cell represent the \emph{p}-value according to this order: ($M_{*},\left\lbrack {Fe/H} \right\rbrack,T_{*} $).}
    \label{tabref:geosciences-343110-t005}

\setlength{\cellWidtha}{\textwidth/4-2\tabcolsep-0.2in}
\setlength{\cellWidthb}{\textwidth/4-2\tabcolsep+0.2in}
\setlength{\cellWidthc}{\textwidth/4-2\tabcolsep+0.2in}
\setlength{\cellWidthd}{\textwidth/4-2\tabcolsep+0.3in}
\scalebox{0.9}[0.9]{\begin{tabular}{>{\PreserveBackslash\centering}m{\cellWidtha}>{\PreserveBackslash\centering}m{\cellWidthb}>{\PreserveBackslash\centering}m{\cellWidthc}>{\PreserveBackslash\centering}m{\cellWidthd}}
\toprule

 & \textbf{\emph{Overlap} OPEN-ARCHIVE} & \textbf{\emph{Overlap} EU-OPEN } & \textbf{\emph{Overlap} ARCHIVE-EU}\\
\cmidrule{1-4}

\textbf{\boldmath{\emph{Unique} EU}} & --- & (\textless{}10\textsuperscript{$-$10}, \textless{}10\textsuperscript{$-$10}, 9 $\times$ 10\textsuperscript{$-$10}) & (5 $\times$ 10\textsuperscript{$-$7}, 0.19, 0.02)\\

\textbf{\boldmath{\emph{Unique} ARCHIVE}} & (\textless{}10\textsuperscript{$-$10}, \textless{}10\textsuperscript{$-$10}, \textless{}10\textsuperscript{$-$10}) & --- & (2 $\times$ 10\textsuperscript{$-$6}, 2 $\times$ 10\textsuperscript{$-$5}, 2 $\times$ 10\textsuperscript{$-$4})\\

\textbf{\boldmath{\emph{Unique} OPEN}} & (0.12, 0.02, 0.09) & (0.06, 0.04, 9 $\times$ 10\textsuperscript{$-$4}) & ---\\

\bottomrule
\end{tabular}}

    \end{table}

Although it would be reasonable to detect most biases in the exoplanets catalogs in the extreme ends of the examined distributions, the analysis we have presented used a quantitative approach to study the biases the catalogs possess. We conclude that the biases in the catalogs are caused by: Some missing or obsolete information about a planets’ properties (e.g., the system of ‘\emph{Kepler 53 b,c}’ is labeled in the ARCHIVE catalog with main planetary masses of $M_{P} $ \textless{} $18~M_{J} $ and $M_{P} $ \textless{} $15~M_{J} $, for planets ‘\emph{b}’, ‘\emph{c}’ respectively~\cite{B25-geosciences-343110}, however a later paper~\cite{B26-geosciences-343110} finds the masses to be much smaller $\left. M_{P} \right.\sim 0.18~M_{J} $, $\left. M_{P} \right.\sim 0.11~M_{J} $, respectively, as listed in both EU and OPEN); model dependent planetary information based on some theoretical assumption and not a direct measurement (e.g., ‘\emph{Kepler-446 b,c,d}’: Both EU and OPEN display their mass property, however, according to the reference article~\cite{B27-geosciences-343110}, the given value is only a coarse estimate for the planets’ masses and expected RV semi-amplitude signatures using recent empirically-measured data); roughly estimated measurements, which is especially relevant for stellar parameters; or approximated upper limits as the nominal value (e.g., ‘\emph{Kepler-114 c}’ is a TTV detected mass planet. The EU catalog displays its mass to be $\left. M_{p} \right.\sim 40M_{\oplus} $~\cite{B26-geosciences-343110}. However, since this measurement is an upper limit, we find it to be inconsistent with the nominal measurement the ARCHIVE displays of $\left. M_{p} \right.\sim 2.8M_{\oplus} $~\cite{B28-geosciences-343110}, while the OPEN catalog does not include this planet).

\section{Summary \& Discussion \label{sect:sec4-geosciences-343110}}

Our analysis suggests that, although the main exoplanet catalogs overlap significantly, which results in similar distributions for most astrophysical parameters, the small discrepancies between the subsets highlight some of the catalogs’ biases. These biases can best be seen in the extreme ends of the examined distributions of small mass, long orbital period planets or small stars (less than \mbox{our sun}). These biases do not only result from different numbers of confirmed planets in each catalog, but mainly from contributing factors, related to the data collection policy of each catalog, such as: \mbox{The process} each catalog uses to present and collect the properties of a specific planet, the decision whether to include a controversial object as a planet, or the routine maintenance each catalog team performs to its current listed~planets.

Furthermore, in our analysis, we excluded planets with masses larger than $M_{p} > 10M_{J} $. However the different catalogs use different mass boundaries, which also adds to their different biases. Unfortunately, most of the biases we found are due to the use of various subjective criteria in compiling and maintaining the database. Although all catalogs usually include in their database planets announced in peer-reviewed publications, this should not be the only criterion for a confirmed planet. We suggest that the explosive growth in the known planet population in recent years once again highlights the need for a more rigorous and objective mechanism to tag planets as confirmed. The differences among the catalogs demonstrate that there are conflicting views in the community regarding such criteria. The International Astronomical Union (IAU) is an objective and well-accepted authority by the community, and we therefore suggest that a central catalog could be maintained by Division F (Planetary Systems and Bio-astronomy) of the IAU, and specifically its Commission F2 (Exoplanets and the Solar Systems). Discussions within the commission should resolve the various differences and arrive at a system that can be agreed~upon.

After performing this analysis and scrutinizing the different calculated biases, we can carefully make the following statements:
\begin{enumerate}[label=$\bullet$]
\item The ARCHIVE catalog is the most up-to-date catalog, with recent Kepler and K2 planet discoveries. It is also the least biased catalog in terms of the interpretation of the mass upper limit, being the true value or the adoption of a model-based value instead of a genuine measurement. Another interesting feature the catalog has is a list of “removed targets” displaying objects that had been listed in the catalog but were removed, suggesting a more rigorous process applied by the ARCHIVE~team.
\item The EU catalog is less restrictive when listing the planetary properties, and therefore could include imprecise estimates. The EU catalog differs the most with the \emph{overlap} subsets, probably due to its more permissive acceptance criteria and the use of mixed sources of information. However, it has the most wide and large coverage of~planets.
\item The OPEN catalog is somewhere in the middle, between ARCHIVE and EU. In some cases, we find that it resembles the EU subsets, while in others the ARCHIVE. This might not be surprising, given that this catalog is an open-source catalog which is managed and updated by the astronomical community. Although its interface is elegant and user friendly, it has its drawbacks, especially the lack of detection reference and a smaller planet~population.
\end{enumerate}

Finally, while each catalog suffers biases, for an exostatistics work, there should not be too much difference among the databases, since the planet population (especially the one compared in \mbox{this work}) is large enough to wash out the small biases and discrepancies. Nevertheless, we find the fusion of catalogs (the \emph{overlap} subset) a powerful tool as a starting point for increasing the reliability of exostatistics research. A promising platform seems to be the Data \& Analysis Center for Exoplanets (DACE) database (\url{https://dace.unige.ch}), which includes a linked table to commonly-used exoplanet catalogs. DACE offers an accessible option to check the properties of a specific planet listed in different catalogs, and to compare its properties as they are displayed on the~catalogs.

Besides a careful and detailed inspection of each exoplanet related paper confirmation, other useful techniques that can be used to increase the confidence of some exoplanet databases is to check other related parameters such as: Discovery date and update times, which can solve issues of “catch-up” times between catalogs and the rate by which they upload new exoplanets; a measure of the velocity semi-amplitude K parameter can suggest the mass measurement is truly deduced from a RV measurement and not derived from some theoretical model; a TTV flag with reported eccentricity parameter can suggest the reported mass measurement is probably not an upper limit, but some nominal~value.

\vspace{6pt}

\authorcontributions{Formal analysis, D.B.; Investigation, D.B., R.H. and S.Z.; Methodology, D.B., R.H. and S.Z.; Supervision, R.H. and S.Z.; Validation, D.B., R.H. and S.Z.; Writing--original draft, D.B.; Writing--review \& editing, D.B., R.H. and S.Z.}
\funding{This research was supported by the Ministry of Science, Technology and Space,~Israel.}
\acknowledgments{We thank the anonymous referees for valuable comments and suggestions. This research has made use of the NASA Exoplanet Archive, which is operated by the California Institute of Technology, under contract with the National Aeronautics and Space Administration under the Exoplanet Exploration Program. This research has made use of the Exoplanet Orbit Database and the Exoplanet Data Explorer at exoplanets.org.

}
\conflictsofinterest{The authors declare no conflict of~interest.}

\appendix

\makeatletter
\@addtoreset{table}{part}
\renewcommand{\thetable}{A\@arabic\c@table}
\setcounter{table}{0}
\@addtoreset{figure}{part}
\renewcommand{\thefigure}{A\@arabic\c@figure}
\setcounter{figure}{0}
\makeatother
\section{\noindent Details of the Comparison \label{app:app1-geosciences-343110}}

\subsection{Comparison of Planetary Properties \label{app:secAdot1-geosciences-343110}}

\begin{figure}[H]
\centering
\includegraphics[scale=1]{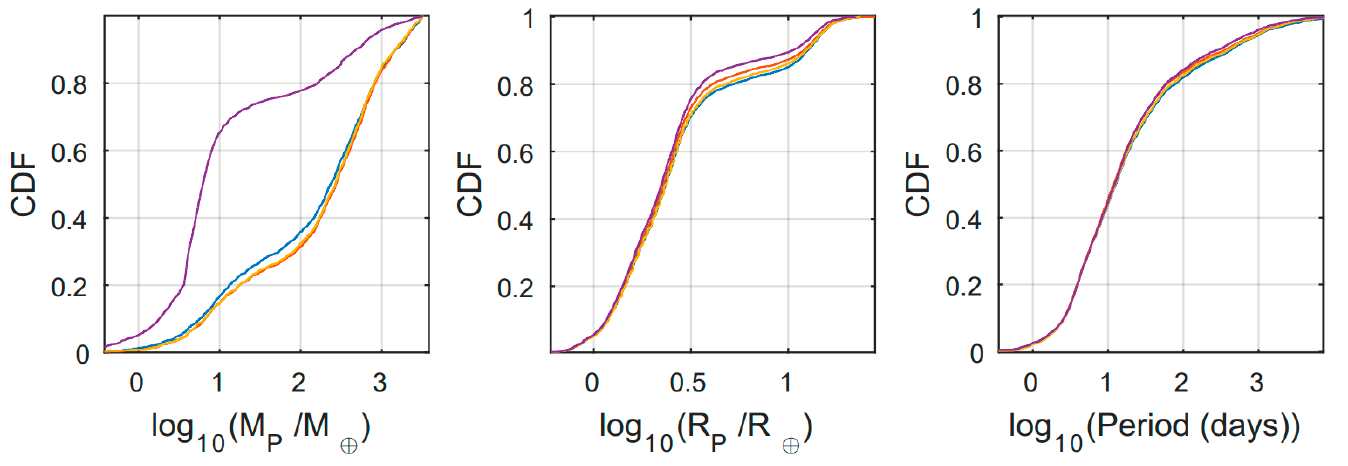}
\caption{Cumulative distribution functions (CDFs) of the planetary mass, radius and orbital period. The different colors correspond to the different catalogs: Blue: The Extrasolar Planets Encyclopaedia (EU), red: NASA Exoplanet Archive (ARCHIVE), orange: Open Exoplanet Catalogue (OPEN), and purple: Exoplanet Data Explorer (ORG).}
\label{fig:geosciences-343110-f0A1}
\end{figure}

We first compared the different catalogs using a KS test for the planet properties of mass ($M_{P} $), radius ($R_{P} $) and orbital period ($Period $). \tabref{tabref:geosciences-343110-t002} summarizes the number of available objects in each catalog for the different properties. We present in \tabref{tabref:geosciences-343110-t0A1} and \fig{fig:geosciences-343110-f0A1} the \emph{p}-values of the corresponding KS tests and relevant empirical cumulative distribution functions (CDF) of each subset respectively. The distributions of the planetary properties from the different catalogs were found to be very similar, apart from the ORG catalog, which bases its missing planetary mass values on a theoretical mass-radius~relation.

    \begin{table}[H]
    \tablesize{\small}
    \centering
    \caption{\emph{p}-values of the different catalogs’ KS test for various planetary~properties.}
    \label{tabref:geosciences-343110-t0A1}

\setlength{\cellWidtha}{\textwidth/7-2\tabcolsep+0.2in}
\setlength{\cellWidthb}{\textwidth/7-2\tabcolsep-0in}
\setlength{\cellWidthc}{\textwidth/7-2\tabcolsep+0.1in}
\setlength{\cellWidthd}{\textwidth/7-2\tabcolsep+0.2in}
\setlength{\cellWidthe}{\textwidth/7-2\tabcolsep-0in}
\setlength{\cellWidthf}{\textwidth/7-2\tabcolsep+0.3in}
\setlength{\cellWidthg}{\textwidth/7-2\tabcolsep-0in}
\scalebox{0.88}[0.88]{\begin{tabular}{>{\PreserveBackslash\centering}m{\cellWidtha}>{\PreserveBackslash\centering}m{\cellWidthb}>{\PreserveBackslash\centering}m{\cellWidthc}>{\PreserveBackslash\centering}m{\cellWidthd}>{\PreserveBackslash\centering}m{\cellWidthe}>{\PreserveBackslash\centering}m{\cellWidthf}>{\PreserveBackslash\centering}m{\cellWidthg}}
\toprule

\textbf{Planet Property} & \textbf{EU-ORG} & \textbf{EU-ARCHIVE} & \textbf{ARCHIVE-ORG} & \textbf{EU-OPEN} & \textbf{OPEN-ARCHIVE} & \textbf{OPEN-ORG}\\
\cmidrule{1-7}

$M_{P} $ & \textless{}10\textsuperscript{$-$10} & 0.12 & \textless{}10\textsuperscript{$-$10} & 0.30 & 0.99 & \textless{}10\textsuperscript{$-$10}\\

$R_{P} $ & 6 $\times$ 10\textsuperscript{$-$4} & 0.21 & 0.18 & 0.82 & 0.03 & 3 $\times$ 10\textsuperscript{$-$4}\\

$Period $ & 0.32 & 0.37 & 0.96 & 0.99 & 0.74 & 0.73\\

\bottomrule
\end{tabular}}

    \end{table}
    \unskip

Although the catalogs seemed similar in this comparison, to reveal their true differences, we applied the analysis discussed in \sect{sect:sec2-geosciences-343110} to compare the \emph{overlap} and \emph{unique} subsets between each pair of the EU-ARCHIVE-OPEN catalogs. In all the comparisons we also investigated whether there was any difference between the properties of the \emph{overlap} parts listed in catalog ‘\emph{A}’ or ‘\emph{B}’. In all cases, the \emph{p}-values were nearly one, suggesting that the two catalogs are comparable. In the following subsections we present the comparison for the different planetary~properties.

\subsubsection{Planetary Mass \label{app:secAdot1dot1-geosciences-343110}}

We summarize the results of the planetary mass comparisons in \cref{tabref:geosciences-343110-t0A2,tabref:geosciences-343110-t0A3}. In these tables (and the following tables), each row represents two catalogs we compare, where the columns represent the number of planets in that subset (\tabref{tabref:geosciences-343110-t0A2}) or the \emph{p}-values of the comparison between each two subsets (\tabref{tabref:geosciences-343110-t0A3}). We abbreviate by ‘\emph{UA}’ and ‘\emph{UB}’ the \emph{unique} subsets ‘\emph{A}’ and ‘\emph{B}’ respectively. \mbox{For instance}, the first row in \tabref{tabref:geosciences-343110-t0A3} represents (by column order from left to right): The \emph{p}-value of the comparison between the \emph{overlap} of the EU and OPEN catalogs with the \emph{unique} part of EU catalog; the \emph{p}-value of the comparison between the \emph{overlap} of the EU and OPEN catalogs with the \emph{unique} part of OPEN catalog; and the \emph{p}-value of the comparison between the \emph{unique} EU and OPEN subsets (see also \fig{fig:geosciences-343110-f001}, for a graphic explanation).

\begin{figure}[H]
\centering
\includegraphics[scale=1]{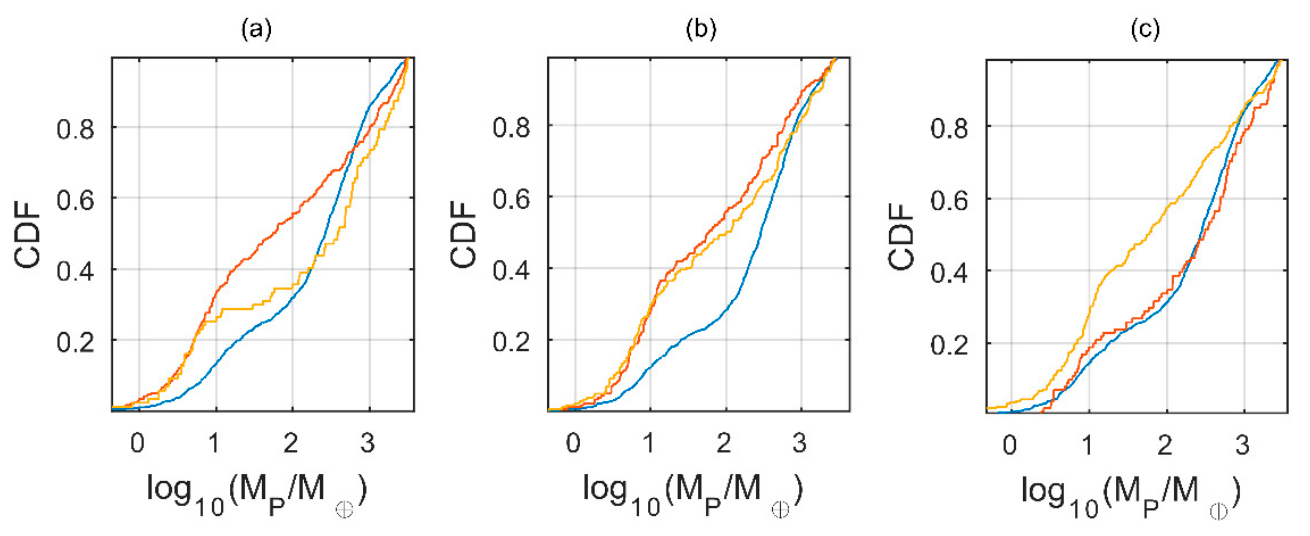}
\caption{\textls[-20]{CDF of the planetary mass property for different subsets. (\textbf{\boldmath{a}}) EU-OPEN; (\textbf{\boldmath{b}}) OPEN-ARCHIVE; (\textbf{\boldmath{c}}) ARCHIVE-EU blue: \emph{Overlap}, red: \emph{UA}, yellow: \emph{UB}.}}
\label{fig:geosciences-343110-f0A2}
\end{figure}

The CDFs of the different subsets is shown in \fig{fig:geosciences-343110-f0A2}. We found that the number of planets in the \emph{overlap} subsets was large, yet the population of the \emph{unique} subsets (in this section and the following to come) were non-negligible. Moreover, although the number of planets in each catalog was different, there was no single catalog that included all the objects from the other catalog. We found that the \emph{unique} subsets were different from the \emph{overlap} subsets with the exception of the \emph{unique} ARCHIVE and OPEN subsets, when comparing them with the \emph{overlap} subset of the EU. Correspondingly, we also found the \emph{unique} parts of the ARCHIVE and OPEN to be very similar based on the comparison between these two~catalogs.

    \begin{table}[H]
    \tablesize{\small}
    \centering
    \caption{Catalogs’ subsets total number of planets with listed planetary~mass.}
    \label{tabref:geosciences-343110-t0A2}

\setlength{\cellWidtha}{\textwidth/4-2\tabcolsep-0.5in}
\setlength{\cellWidthb}{\textwidth/4-2\tabcolsep-0.9in}
\setlength{\cellWidthc}{\textwidth/4-2\tabcolsep-1in}
\setlength{\cellWidthd}{\textwidth/4-2\tabcolsep-1in}
\scalebox{1}[1]{\begin{tabular}{>{\PreserveBackslash\centering}m{\cellWidtha}>{\PreserveBackslash\centering}m{\cellWidthb}>{\PreserveBackslash\centering}m{\cellWidthc}>{\PreserveBackslash\centering}m{\cellWidthd}}
\toprule

\textbf{Catalog} & \textbf{\emph{Overlap}} & \textbf{\emph{UA}} & \textbf{\emph{UB}}\\
\cmidrule{1-4}

EU-OPEN & 1188 & 239 & 87\\

OPEN-ARCHIVE & 1089 & 187 & 186\\

ARCHIVE-EU & 1175 & 101 & 252\\

\bottomrule
\end{tabular}}

    \end{table}
    \unskip
    
    \begin{table}[H]
    \tablesize{\small}
    \centering
    \caption{\emph{p}-values of the different catalogs’ KS test for the planetary mass~property.}
    \label{tabref:geosciences-343110-t0A3}

\setlength{\cellWidtha}{\textwidth/4-2\tabcolsep-0.3in}
\setlength{\cellWidthb}{\textwidth/4-2\tabcolsep-0.5in}
\setlength{\cellWidthc}{\textwidth/4-2\tabcolsep-0.5in}
\setlength{\cellWidthd}{\textwidth/4-2\tabcolsep-0.6in}
\scalebox{1}[1]{\begin{tabular}{>{\PreserveBackslash\centering}m{\cellWidtha}>{\PreserveBackslash\centering}m{\cellWidthb}>{\PreserveBackslash\centering}m{\cellWidthc}>{\PreserveBackslash\centering}m{\cellWidthd}}
\toprule

\textbf{Catalog} & \textbf{\emph{Overlap} vs. \emph{UA}} & \textbf{\emph{Overlap} vs. \emph{UB}} & \textbf{\emph{UA} vs. \emph{UB}}\\
\cmidrule{1-4}

EU-OPEN  & \textless{}10\textsuperscript{$-$10} & 0.034 & 7 $\times$ 10\textsuperscript{$-$3}\\

OPEN-ARCHIVE  & \textless{}10\textsuperscript{$-$10} & 3 $\times$ 10\textsuperscript{$-$8} & 0.42\\

ARCHIVE-EU & 0.69 & \textless{}10\textsuperscript{$-$10} & 3.5 $\times$ 10\textsuperscript{$-$4}\\

\bottomrule
\end{tabular}}

    \end{table}
    \unskip

We found that most of the disagreement between the \emph{overlap} and \emph{unique} subsets to be concentrated in the region of lower-mass planets. At present, most planets with masses $< 10M_{\oplus} $, are estimated from TTVs in multi-planetary systems. Usually, the mass of a planet detected through TTV is not resolved directly, and in practice, is degenerate with the planet’s eccentric orbit~\cite{B18-geosciences-343110}, following knowledge on an upper mass limit, rather than a nominal value. Consequently, we found the catalogs use different inclusion criteria for these kinds of planets with no uniform approach to displaying this mass limit. Sometimes the catalogs chose to omit the value altogether, sometimes to display it as an upper limit, but in many cases, the mass limit was reported as a legitimate nominal estimate. The derived CDF presented in \fig{fig:geosciences-343110-f0A2} suggest the EU catalog, as opposed to the ARCHIVE and OPEN, displays many of its TTV planets’ mass as the reported upper limit value given from their confirmation paper without any strict filtering. The modest agreement we found between the OPEN and ARCHIVE \emph{unique} parts was artificial, driven from the similar number of TTV planets the two catalogs choose to~include.

\subsubsection{Planetary Radius and Orbital Period \label{app:secAdot1dot2-geosciences-343110}}

\begin{figure}[H]
\centering
\includegraphics[scale=1]{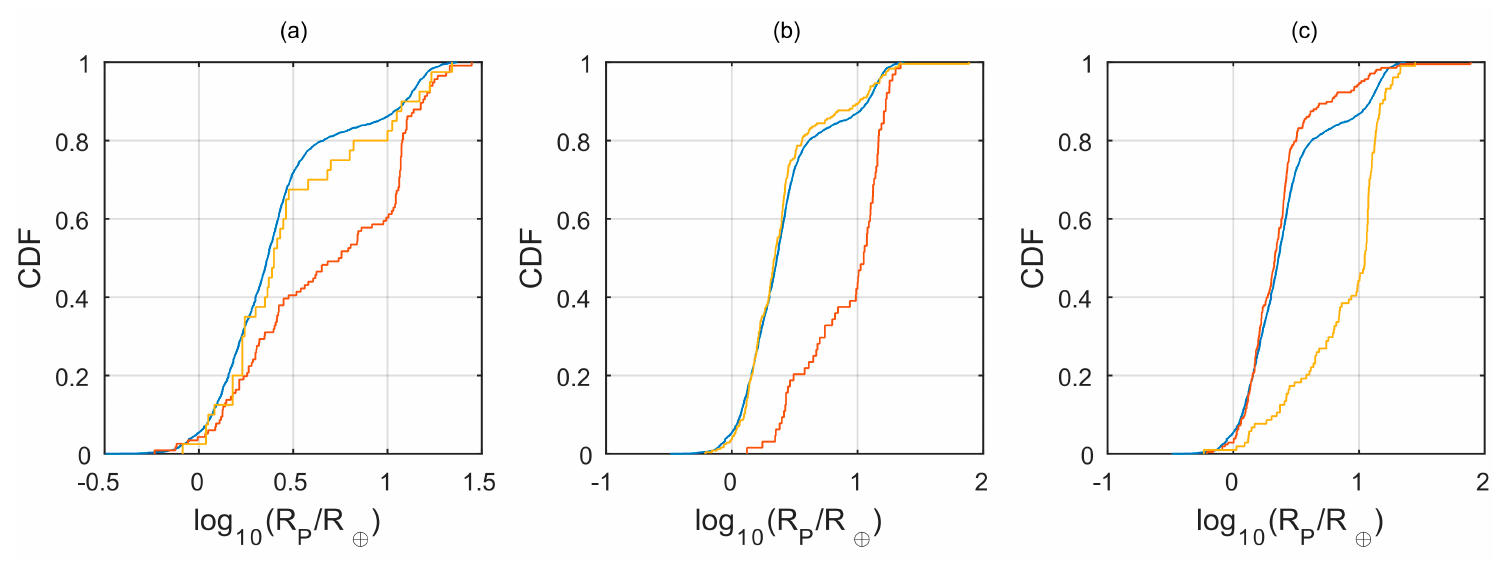}
\caption{\textls[-20]{CDF of the planetary radius property for different subsets. (\textbf{\boldmath{a}}) EU-OPEN; (\textbf{\boldmath{b}}) OPEN-ARCHIVE; (\textbf{\boldmath{c}}) ARCHIVE-EU blue: \emph{Overlap}, red: \emph{UA}, yellow: \emph{UB}.}}
\label{fig:geosciences-343110-f0A3}
\end{figure}

\begin{figure}[H]
\centering
\includegraphics[scale=1]{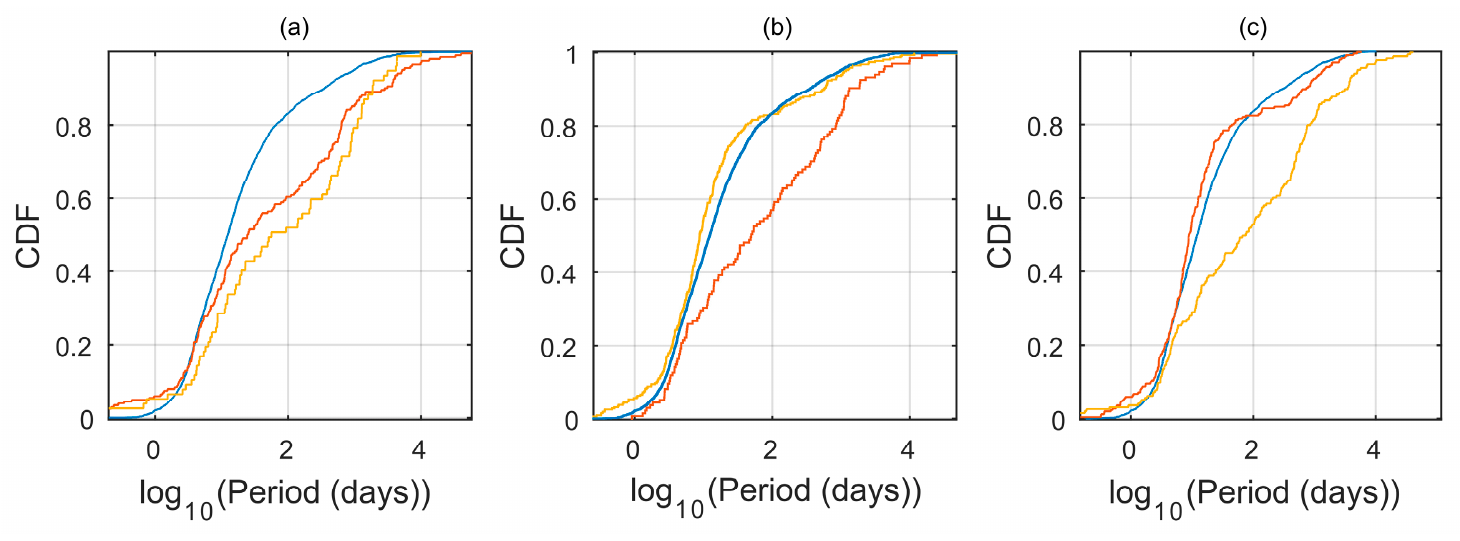}
\caption{CDF of the planetary orbital period property for different subsets. (\textbf{\boldmath{a}}) EU-OPEN; (\textbf{\boldmath{b}}) OPEN-ARCHIVE; (\textbf{\boldmath{c}}) ARCHIVE-EU blue: \emph{Overlap}, red: \emph{UA}, yellow: \emph{UB}.}
\label{fig:geosciences-343110-f0A4}
\end{figure}

Although we inspected the properties of planetary radius and orbital period separately, we found the result of the comparison analysis between their \emph{overlap} and \emph{unique} subsets to be similar. We present the number of planets with reported radii, calculated \emph{p}-values and CDFs of the different subsets in \cref{tabref:geosciences-343110-t0A4,tabref:geosciences-343110-t0A5} and \fig{fig:geosciences-343110-f0A3}, respectively. We present the number of planetary orbital periods and calculated \emph{p}-values and CDFs of the different subsets in \cref{tabref:geosciences-343110-t0A6,tabref:geosciences-343110-t0A7} and \fig{fig:geosciences-343110-f0A4},~respectively.

    \begin{table}[H]
    \tablesize{\small}
    \centering
    \caption{Catalogs’ subsets total number of planets with listed planetary~radius.}
    \label{tabref:geosciences-343110-t0A4}

\setlength{\cellWidtha}{\textwidth/4-2\tabcolsep-0.5in}
\setlength{\cellWidthb}{\textwidth/4-2\tabcolsep-0.9in}
\setlength{\cellWidthc}{\textwidth/4-2\tabcolsep-1in}
\setlength{\cellWidthd}{\textwidth/4-2\tabcolsep-1in}
\scalebox{1}[1]{\begin{tabular}{>{\PreserveBackslash\centering}m{\cellWidtha}>{\PreserveBackslash\centering}m{\cellWidthb}>{\PreserveBackslash\centering}m{\cellWidthc}>{\PreserveBackslash\centering}m{\cellWidthd}}
\toprule

\textbf{Catalog} & \textbf{\emph{Overlap}} & \textbf{\emph{UA}} & \textbf{\emph{UB}}\\
\cmidrule{1-4}

EU-OPEN  & 2691 & 116 & 40\\

OPEN-ARCHIVE  & 2667 & 64 & 244\\

ARCHIVE-EU & 2703 & 208 & 104\\

\bottomrule
\end{tabular}}

    \end{table}
    \unskip
    
    \begin{table}[H]
    \tablesize{\small}
    \centering
    \caption{\emph{p}-values of the different catalogs’ KS test for the planetary radius~property.}
    \label{tabref:geosciences-343110-t0A5}

\setlength{\cellWidtha}{\textwidth/4-2\tabcolsep-0.3in}
\setlength{\cellWidthb}{\textwidth/4-2\tabcolsep-0.5in}
\setlength{\cellWidthc}{\textwidth/4-2\tabcolsep-0.5in}
\setlength{\cellWidthd}{\textwidth/4-2\tabcolsep-0.6in}
\scalebox{1}[1]{\begin{tabular}{>{\PreserveBackslash\centering}m{\cellWidtha}>{\PreserveBackslash\centering}m{\cellWidthb}>{\PreserveBackslash\centering}m{\cellWidthc}>{\PreserveBackslash\centering}m{\cellWidthd}}
\toprule

\textbf{Catalog} & \textbf{\emph{Overlap} vs. \emph{UA}} & \textbf{\emph{Overlap} vs. \emph{UB}} & \textbf{\emph{UA} vs. \emph{UB}}\\
\cmidrule{1-4}

EU-OPEN & \textless{}10\textsuperscript{$-$10} & 0.74 & 0.02\\

OPEN-ARCHIVE & \textless{}10\textsuperscript{$-$10} & 0.03 & \textless{}10\textsuperscript{$-$10}\\

ARCHIVE-EU & 0.01 & \textless{}10\textsuperscript{$-$10} & \textless{}10\textsuperscript{$-$10}\\

\bottomrule
\end{tabular}}

    \end{table}
    \unskip

\unskip
    
    \begin{table}[H]
    \tablesize{\small}
    \centering
    \caption{Catalogs’ subsets total number of planets with listed planetary orbital~period.}
    \label{tabref:geosciences-343110-t0A6}

\setlength{\cellWidtha}{\textwidth/4-2\tabcolsep-0.5in}
\setlength{\cellWidthb}{\textwidth/4-2\tabcolsep-0.9in}
\setlength{\cellWidthc}{\textwidth/4-2\tabcolsep-1in}
\setlength{\cellWidthd}{\textwidth/4-2\tabcolsep-1in}
\scalebox{1}[1]{\begin{tabular}{>{\PreserveBackslash\centering}m{\cellWidtha}>{\PreserveBackslash\centering}m{\cellWidthb}>{\PreserveBackslash\centering}m{\cellWidthc}>{\PreserveBackslash\centering}m{\cellWidthd}}
\toprule

\textbf{Catalog} & \textbf{\emph{Overlap}} & \textbf{\emph{UA}} & \textbf{\emph{UB}}\\
\cmidrule{1-4}

EU-OPEN & 3303 & 202 & 77\\

OPEN-ARCHIVE & 3245 & 135 & 317\\

ARCHIVE-EU & 3313 & 249 & 192\\

\bottomrule
\end{tabular}}

    \end{table}
    \unskip
    
    \begin{table}[H]
    \tablesize{\small}
    \centering
    \caption{\emph{p}-values of the different catalogs’ KS test for the planetary orbital period~property.}
    \label{tabref:geosciences-343110-t0A7}

\setlength{\cellWidtha}{\textwidth/4-2\tabcolsep-0.3in}
\setlength{\cellWidthb}{\textwidth/4-2\tabcolsep-0.5in}
\setlength{\cellWidthc}{\textwidth/4-2\tabcolsep-0.5in}
\setlength{\cellWidthd}{\textwidth/4-2\tabcolsep-0.6in}
\scalebox{1}[1]{\begin{tabular}{>{\PreserveBackslash\centering}m{\cellWidtha}>{\PreserveBackslash\centering}m{\cellWidthb}>{\PreserveBackslash\centering}m{\cellWidthc}>{\PreserveBackslash\centering}m{\cellWidthd}}
\toprule

\textbf{Catalog} & \textbf{\emph{Overlap} vs. \emph{UA}} & \textbf{\emph{Overlap} vs. \emph{UB}} & \textbf{\emph{UA} vs. \emph{UB}}\\
\cmidrule{1-4}

EU-OPEN  & \textless{}10\textsuperscript{$-$10} & 4 $\times$ 10\textsuperscript{$-$8} & 0.31\\

OPEN-ARCHIVE  & 10\textsuperscript{$-$9} & 2 $\times$ 10\textsuperscript{$-$3} & 5 $\times$ 10\textsuperscript{$-$10}\\

ARCHIVE-EU & 0.02 & \textless{}10\textsuperscript{$-$10} & \textless{}10\textsuperscript{$-$10}\\

\bottomrule
\end{tabular}}

    \end{table}
    \unskip

In the analysis of both properties, we found relative similarity between the \emph{overlap} and \emph{unique} subsets of the ARCHIVE catalog and a large difference with the EU and OPEN subsets. It seems that the OPEN and EU \emph{unique} parts are shifted towards larger radii ($R_{P} > 1R_{J} $) or longer orbital periods. However, when we examined the list of ARCHIVE \emph{unique} planets we found over 75\% of them to be K2 planets. We found also a similar ratio of Kepler and K2 planets that populate the \emph{overlap} ARCHIVE subset. These two facts together suggest the reason for the good agreement we found between the ARCHIVE \emph{overlap} and \emph{unique} subsets related to the similar properties and biases that emerged from the transiting detected Kepler and K2 planets: Relative small radii ($R_{P} < 4R_{\oplus} $) and short orbital period ($Period < 100$~days)~\cite{B29-geosciences-343110}. Other differences we found between the listed subsets are: Listed planets where the confirmation paper has used some theoretical M--R relations to infer the planetary radius (or mass), some unusual large radii planets suffering strong tidal forces due to their proximity to the parent stars (in short periods), planets with a ‘strange’ transiting light curves that make the planetary detection more controversial,~etc.

We conclude that the reason for the disagreement between the catalogs radii and orbital period distributions is derived from: (1) Different criteria for including planets in the catalogs that do not emerge directly from the criteria presented in \tabref{tabref:geosciences-343110-t001}; (2) Different update and confirmation processes for new~candidates.

\subsubsection{Planetary Mass-Radius-Period (MRP) \label{app:secAdot1dot3-geosciences-343110}}

\begin{figure}[H]
\centering
\includegraphics[scale=1]{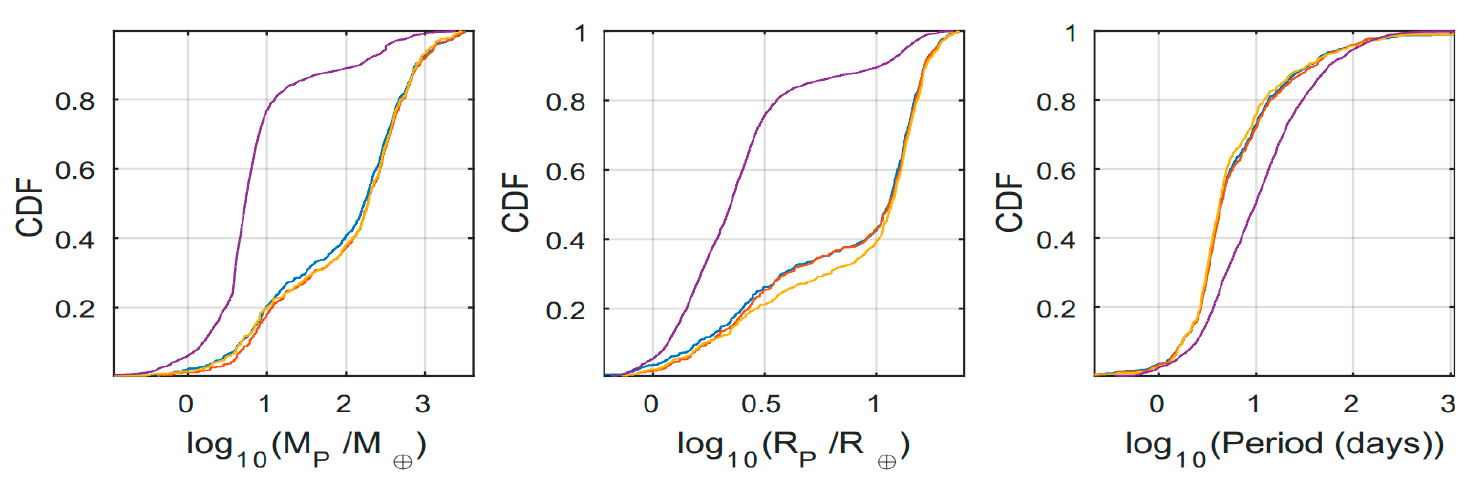}
\caption{Same as \fig{fig:geosciences-343110-f0A1}, but for planets with combined information of planetary~properties.}
\label{fig:geosciences-343110-f0A5}
\end{figure}

In this subsection we compare the total number of planets in each catalog for a subset in which all three properties of planetary mass, radius and orbital period are available. Information of these three physical properties can provide important constraints for planetary characterization and formation models~\cite{B7-geosciences-343110}. Most of the current MRP planets were detected with both RV and transit methods while, as of today, most confirmed mass-radius planets are not TTV detected~\cite{B30-geosciences-343110}. Consequently, we expect to detect especially large mass-radius, close orbit planets (Hot-Jupiters) around bright solar-like stars. As before, we started by analyzing the MRP distributions of the different catalogs (including the \mbox{ORG catalog}). We present in \tabref{tabref:geosciences-343110-t0A8} and \fig{fig:geosciences-343110-f0A5} the \emph{p}-values of the corresponding KS tests and relevant empirical cumulative distribution functions (CDF) of each subset~respectively.

    \begin{table}[H]
    \tablesize{\small}
    \centering
    \caption{Same as \tabref{tabref:geosciences-343110-t0A1}, but for planets with combined information of planetary~properties.}
    \label{tabref:geosciences-343110-t0A8}

\setlength{\cellWidtha}{\textwidth/7-2\tabcolsep+0.2in}
\setlength{\cellWidthb}{\textwidth/7-2\tabcolsep-0in}
\setlength{\cellWidthc}{\textwidth/7-2\tabcolsep+0.1in}
\setlength{\cellWidthd}{\textwidth/7-2\tabcolsep+0.2in}
\setlength{\cellWidthe}{\textwidth/7-2\tabcolsep-0in}
\setlength{\cellWidthf}{\textwidth/7-2\tabcolsep+0.3in}
\setlength{\cellWidthg}{\textwidth/7-2\tabcolsep-0in}
\scalebox{0.88}[0.88]{\begin{tabular}{>{\PreserveBackslash\centering}m{\cellWidtha}>{\PreserveBackslash\centering}m{\cellWidthb}>{\PreserveBackslash\centering}m{\cellWidthc}>{\PreserveBackslash\centering}m{\cellWidthd}>{\PreserveBackslash\centering}m{\cellWidthe}>{\PreserveBackslash\centering}m{\cellWidthf}>{\PreserveBackslash\centering}m{\cellWidthg}}
\toprule

\textbf{Planet Property} & \textbf{EU-ORG} & \textbf{EU-ARCHIVE} & \textbf{ARCHIVE-ORG} & \textbf{EU-OPEN} & \textbf{OPEN-ARCHIVE} & \textbf{OPEN-ORG}\\
\cmidrule{1-7}

$M_{P} $ & \textless{}10\textsuperscript{$-$10} & 0.73 & \textless{}10\textsuperscript{$-$10} & 0.69 & 0.92 & \textless{}10\textsuperscript{$-$10}\\

$R_{P} $ & \textless{}10\textsuperscript{$-$10} & 0.98 & \textless{}10\textsuperscript{$-$10} & 0.21 & 0.24 & \textless{}10\textsuperscript{$-$10}\\

$Period $ & \textless{}10\textsuperscript{$-$10} & 0.99 & \textless{}10\textsuperscript{$-$10} & 0.81 & 0.58 & \textless{}10\textsuperscript{$-$10}\\

\bottomrule
\end{tabular}}

    \end{table}
    \unskip

Similar with the overall catalogs comparison, excluding the ORG catalog with its added theoretical supplement, all three other catalogs agreed on their distributions. We noted that as expected, the distributions were different than those displayed for the separate properties (see \fig{fig:geosciences-343110-f0A1}), with indeed a higher number of planets in the large mass-radius and small period~regimes.

\begin{figure}[H]
\centering
\includegraphics[scale=1]{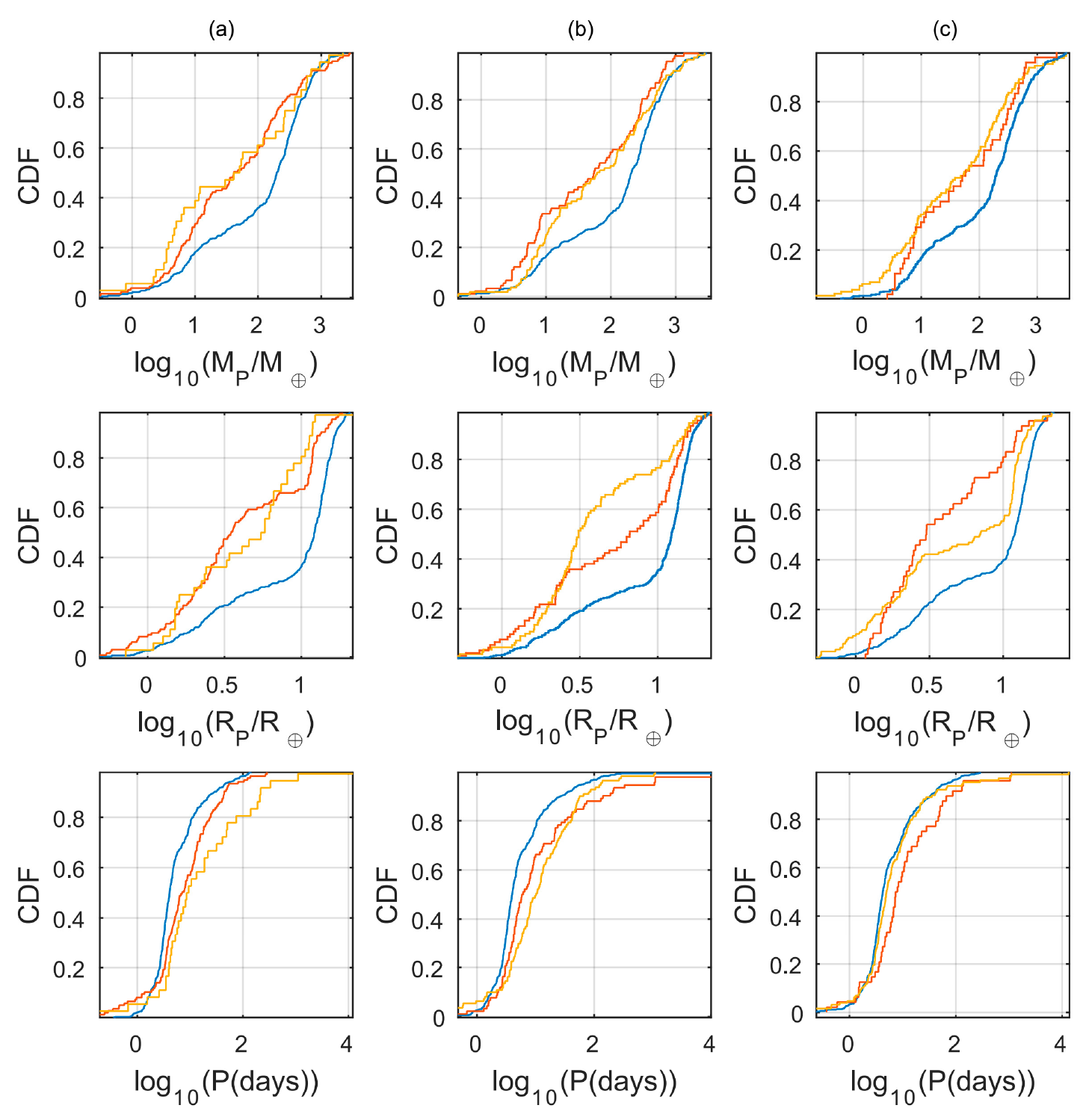}
\caption{CDF of the MRP properties for different subsets. (\textbf{\boldmath{a}}) EU-OPEN; (\textbf{\boldmath{b}}) OPEN-ARCHIVE; (\textbf{\boldmath{c}}) ARCHIVE-EU blue: \emph{Overlap}, red: \emph{UA}, yellow: \emph{UB.}}
\label{fig:geosciences-343110-f0A6}
\end{figure}

By comparing the total numbers, \emph{p}-values and distributions of the catalogs subsets (\cref{tabref:geosciences-343110-t0A9,tabref:geosciences-343110-t0A10} and \fig{fig:geosciences-343110-f0A6}), we found most of the disagreements between the \emph{overlap} and \emph{unique} subsets to be evident in the small planetary regime of TTVs’ detected planets. We found the \emph{unique} subsets, and especially the planetary masses distributions, to be similar between the catalogs, although the listed planets are different. We conclude that most of the disagreement with the planets of listed MRP comes especially from the different approaches each catalog uses for the confirmed TTV~planets.

    \begin{table}[H]
    \tablesize{\small}
    \centering
    \caption{Catalogs’ subsets total number of planets with listed mass-radius-period (MRP)~properties.}
    \label{tabref:geosciences-343110-t0A9}

\setlength{\cellWidtha}{\textwidth/4-2\tabcolsep-0.5in}
\setlength{\cellWidthb}{\textwidth/4-2\tabcolsep-0.9in}
\setlength{\cellWidthc}{\textwidth/4-2\tabcolsep-1in}
\setlength{\cellWidthd}{\textwidth/4-2\tabcolsep-1in}
\scalebox{1}[1]{\begin{tabular}{>{\PreserveBackslash\centering}m{\cellWidtha}>{\PreserveBackslash\centering}m{\cellWidthb}>{\PreserveBackslash\centering}m{\cellWidthc}>{\PreserveBackslash\centering}m{\cellWidthd}}
\toprule

\textbf{Catalog} & \textbf{\emph{Overlap}} & \textbf{\emph{UA}} & \textbf{\emph{UB}}\\
\cmidrule{1-4}

EU-OPEN & 520 & 135 & 36\\

OPEN-ARCHIVE & 465 & 91 & 110\\

ARCHIVE-EU & 528 & 47 & 127\\

\bottomrule
\end{tabular}}

    \end{table}
    \unskip
    
    \begin{table}[H]
    \tablesize{\small}
    \centering
    \caption{\emph{p}-values of the different catalogs’ KS test for the planetary MRP~properties.}
    \label{tabref:geosciences-343110-t0A10}

\setlength{\cellWidtha}{\textwidth/5-2\tabcolsep-0in}
\setlength{\cellWidthb}{\textwidth/5-2\tabcolsep-0in}
\setlength{\cellWidthc}{\textwidth/5-2\tabcolsep-0.1in}
\setlength{\cellWidthd}{\textwidth/5-2\tabcolsep-0.2in}
\setlength{\cellWidthe}{\textwidth/5-2\tabcolsep-0.2in}
\scalebox{1}[1]{\begin{tabular}{>{\PreserveBackslash\centering}m{\cellWidtha}>{\PreserveBackslash\centering}m{\cellWidthb}>{\PreserveBackslash\centering}m{\cellWidthc}>{\PreserveBackslash\centering}m{\cellWidthd}>{\PreserveBackslash\centering}m{\cellWidthe}}
\toprule

\textbf{Catalog} & \textbf{Planetary Property} & \textbf{\emph{Overlap} vs. \emph{UA}} & \textbf{\emph{Overlap} vs. \emph{UB}} & \textbf{\emph{UA} vs. \emph{UB}}\\
\cmidrule{1-5}

  & $M_{P} $ & 2 $\times$ 10\textsuperscript{$-$8} & 6 $\times$ 10\textsuperscript{$-$3} & 0.49\\

EU-OPEN & $R_{P} $ & \textless{}10\textsuperscript{$-$10} & 2 $\times$ 10\textsuperscript{$-$7} & 0.46\\

  & $Period $ & 8 $\times$ 10\textsuperscript{$-$6} & 10\textsuperscript{$-$3} & 0.23\\
\cmidrule{1-5}
  & $M_{P} $ & 6 $\times$ 10\textsuperscript{$-$5} & 3 $\times$ 10\textsuperscript{$-$5} & 0.34\\

OPEN-ARCHIVE  & $R_{P} $ & 7 $\times$ 10\textsuperscript{$-$5} & \textless{}10\textsuperscript{$-$10} & 2 $\times$ 10\textsuperscript{$-$3}\\

  & $Period $ & 6 $\times$ 10\textsuperscript{$-$4} & 3 $\times$ 10\textsuperscript{$-$11} & 0.05\\
\cmidrule{1-5}
  & $M_{P} $ & 0.02 & 3 $\times$ 10\textsuperscript{$-$7} & 0.51\\

ARCHIVE-EU & $R_{P} $ & 3 $\times$ 10\textsuperscript{$-$7} & 8 $\times$ 10\textsuperscript{$-$7} & 0.02\\

  & $Period $ & 8 $\times$ 10\textsuperscript{$-$4} & 0.23 & 0.05\\

\bottomrule
\end{tabular}}

    \end{table}
    \unskip

\unskip

\subsection{Planetary Systems \label{app:secAdot2-geosciences-343110}}

\begin{figure}[H]
\centering
\includegraphics[scale=1]{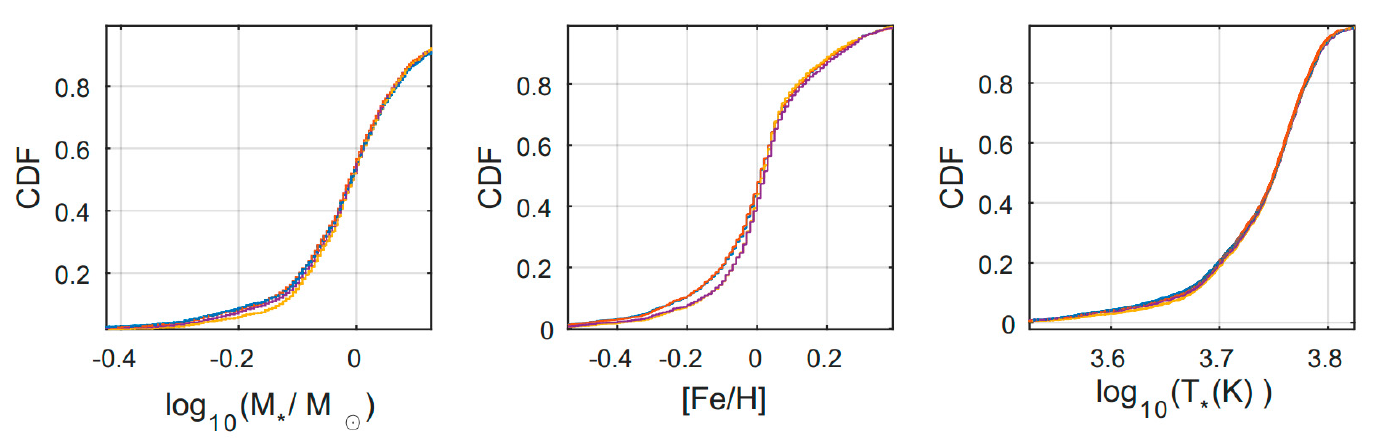}
\caption{CDFs of the stellar mass, metallicity and surface temperature. The different colors correspond to the different catalogs: blue: EU, red: ARCHIVE, orange: OPEN purple:~ORG.}
\label{fig:geosciences-343110-f0A7}
\end{figure}

We first compared the different catalogs by using a KS test for the stellar properties of the planetary systems of mass ($M_{*} $), metallicity ($\lbrack Fe/H $]) and surface temperature ($T_{eff} $). \tabref{tabref:geosciences-343110-t002} summarizes the number of available objects in each catalog for the different properties. We present in \tabref{tabref:geosciences-343110-t0A11} and \fig{fig:geosciences-343110-f0A7} the \emph{p}-values of the corresponding KS tests and relevant empirical CDFs of each subset respectively. We found the distributions to be very similar between all catalogs. The slightly lower metallicity \emph{p}-values we found was caused by the large error that follows the metallicity property measurement and does not relate to any different distribution between the~catalogs.

    \begin{table}[H]
    \tablesize{\small}
    \centering
    \caption{\emph{p}-values of the different catalogs’ KS test for various stellar~properties.}
    \label{tabref:geosciences-343110-t0A11}

\setlength{\cellWidtha}{\textwidth/7-2\tabcolsep+0.2in}
\setlength{\cellWidthb}{\textwidth/7-2\tabcolsep-0in}
\setlength{\cellWidthc}{\textwidth/7-2\tabcolsep+0.1in}
\setlength{\cellWidthd}{\textwidth/7-2\tabcolsep+0.2in}
\setlength{\cellWidthe}{\textwidth/7-2\tabcolsep-0in}
\setlength{\cellWidthf}{\textwidth/7-2\tabcolsep+0.3in}
\setlength{\cellWidthg}{\textwidth/7-2\tabcolsep-0in}
\scalebox{0.88}[0.88]{\begin{tabular}{>{\PreserveBackslash\centering}m{\cellWidtha}>{\PreserveBackslash\centering}m{\cellWidthb}>{\PreserveBackslash\centering}m{\cellWidthc}>{\PreserveBackslash\centering}m{\cellWidthd}>{\PreserveBackslash\centering}m{\cellWidthe}>{\PreserveBackslash\centering}m{\cellWidthf}>{\PreserveBackslash\centering}m{\cellWidthg}}
\toprule

\textbf{Stellar Property} & \textbf{EU-ORG} & \textbf{EU-ARCHIVE} & \textbf{ARCHIVE-ORG} & \textbf{EU-OPEN} & \textbf{OPEN-ARCHIVE} & \textbf{OPEN-ORG}\\
\cmidrule{1-7}

$M_{*} $ & 0.77 & 0.59 & 0.56 & 0.76 & 0.33 & 0.93\\

$\left\lbrack {Fe/H} \right\rbrack $ & 0.08 & 0.99 & 0.03 & 0.07 & 0.01 & 0.28\\

$T_{eff} $ & 0.55 & 0.99 & 0.74 & 0.94 & 0.89 & 0.99\\

\bottomrule
\end{tabular}}

    \end{table}
    \unskip

Similar with the planetary properties, we next compared the \emph{overlap} and \emph{unique} stellar properties subsets between each pair of the EU-ARCHIVE-OPEN catalogs. We performed the following analysis for the stars of each planetary system. As mentioned in \sect{sect:sec2-geosciences-343110}, we also compared the distributions of the first detect planet in each system to better understand the reasons for the possible~differences.

\subsubsection{Stellar Mass and Surface Temperature \label{app:secAdot2dot1-geosciences-343110}}

\begin{figure}[H]
\centering
\includegraphics[scale=1]{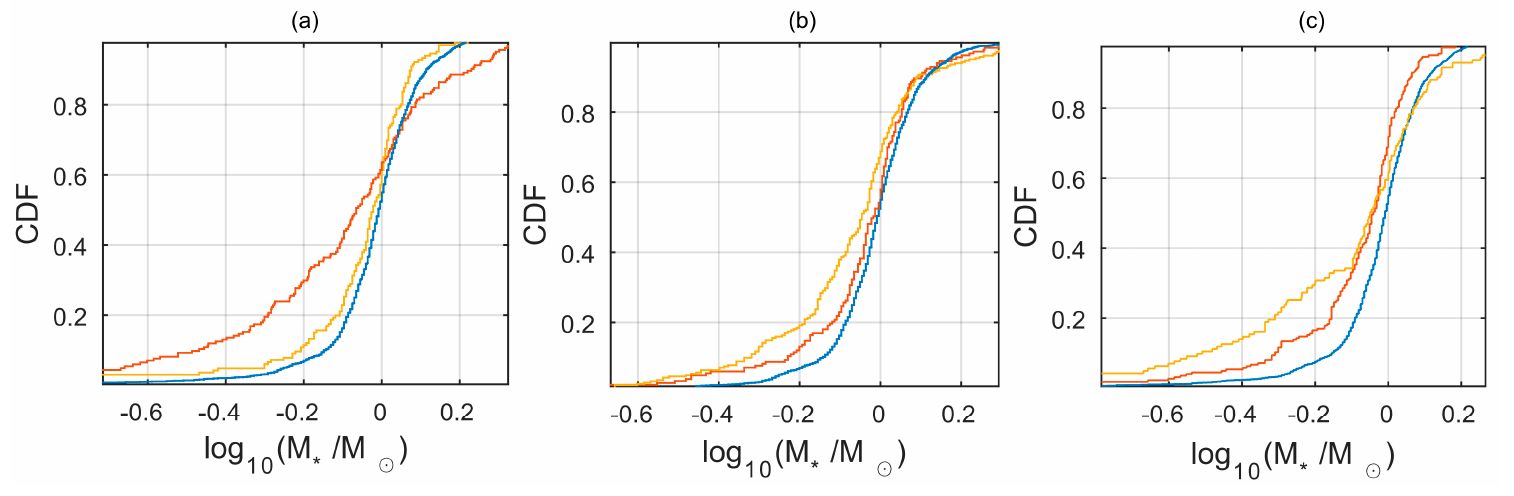}
\caption{CDF of the stellar mass property for different subsets. (\textbf{\boldmath{a}}) EU-OPEN; (\textbf{\boldmath{b}}) OPEN-ARCHIVE; (\textbf{\boldmath{c}}) ARCHIVE-EU blue: \emph{Overlap}, red: \emph{UA}, yellow: \emph{UB.}}
\label{fig:geosciences-343110-f0A8}
\end{figure}

\begin{figure}[H]
\centering
\includegraphics[scale=0.9]{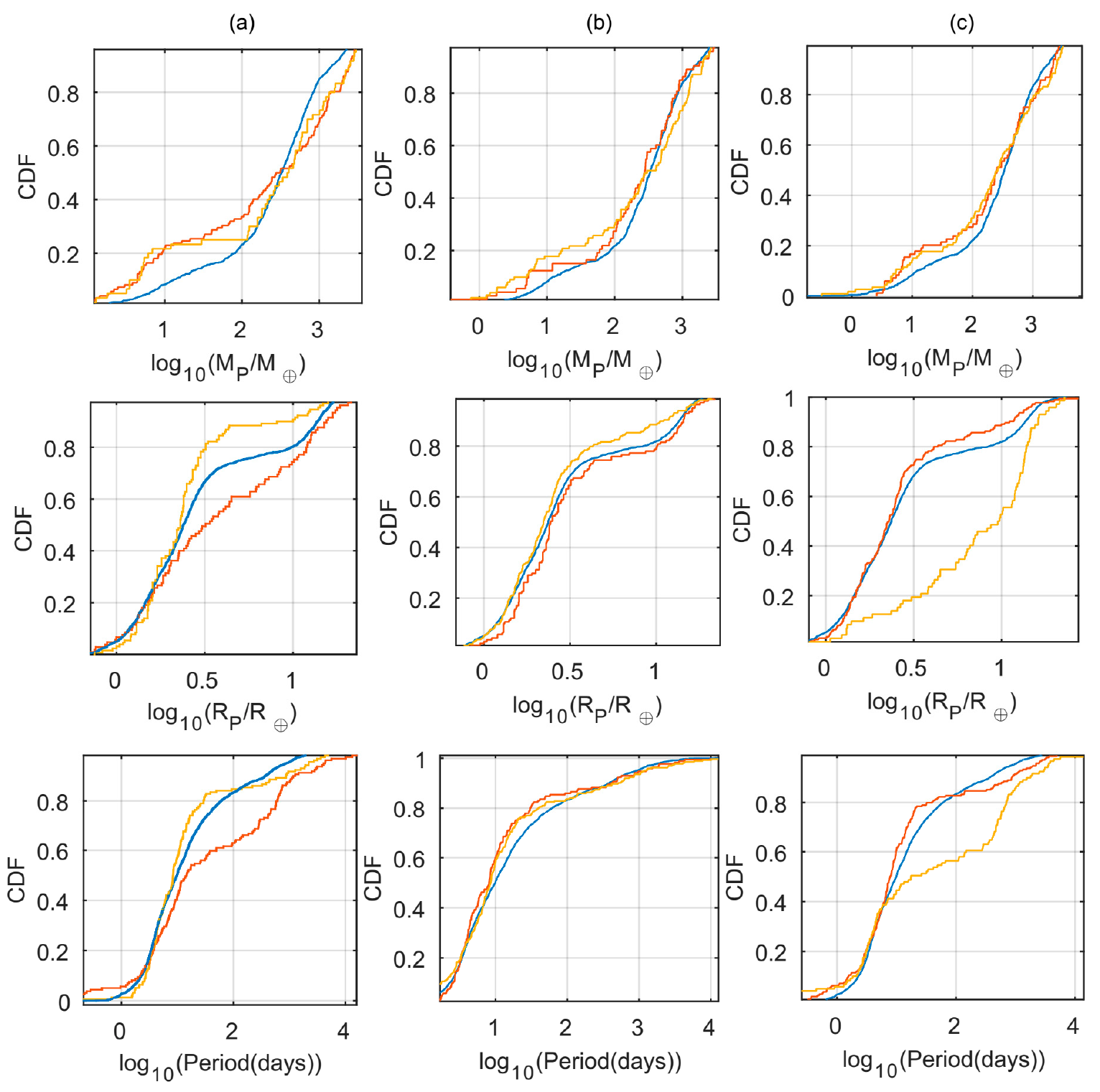}
\caption{CDF of the planetary properties for different subsets of the first detected planet systems with reported stellar mass property. (\textbf{\boldmath{a}}) EU-OPEN; (\textbf{\boldmath{b}}) OPEN-ARCHIVE; (\textbf{\boldmath{c}}) ARCHIVE-EU blue: \emph{Overlap}, red: \emph{UA}, yellow: \emph{UB}.}
\label{fig:geosciences-343110-f0A9}
\end{figure}

Finding the stellar mass is an important prior when assessing a planet’s mass by the RV method~\cite{B31-geosciences-343110}. We expected to detect most of our \emph{overlap} planets around F, G and K stars since most of the planetary detection projects and efforts as of today have been dedicated towards searching planets around a solar mass star. We present the number of objects with reported stellar mass (and first planet’s planetary properties), calculated \emph{p}-values and CDFs of the different subsets in \cref{tabref:geosciences-343110-t0A12,tabref:geosciences-343110-t0A13} and \cref{fig:geosciences-343110-f0A8,fig:geosciences-343110-f0A9}, respectively. We found the \emph{unique} vs \emph{overlap} subsets to be different, with higher disagreement in the regime of small-mass stars (especially K and M-stars). The spectrum of M-stars presents a difficulty for measuring, due to its intrinsic faintness and high activity~\cite{B20-geosciences-343110}. Consequently, detection of planets around these stars is supposed to be more difficult and somewhat controversial, thus explaining why we found a higher relative number of these objects in the \emph{unique}~subsets.

    \begin{table}[H]
    \tablesize{\small}
    \centering
    \caption{Catalogs’ subsets total number of objects with listed stellar mass and planetary properties~fractions.}
    \label{tabref:geosciences-343110-t0A12}

\setlength{\cellWidtha}{\textwidth/5-2\tabcolsep-0in}
\setlength{\cellWidthb}{\textwidth/5-2\tabcolsep+0.4in}
\setlength{\cellWidthc}{\textwidth/5-2\tabcolsep-0.6in}
\setlength{\cellWidthd}{\textwidth/5-2\tabcolsep-0.8in}
\setlength{\cellWidthe}{\textwidth/5-2\tabcolsep-0.6in}
\scalebox{1}[1]{\begin{tabular}{>{\PreserveBackslash\centering}m{\cellWidtha}>{\PreserveBackslash\centering}m{\cellWidthb}>{\PreserveBackslash\centering}m{\cellWidthc}>{\PreserveBackslash\centering}m{\cellWidthd}>{\PreserveBackslash\centering}m{\cellWidthe}}
\toprule

\textbf{Catalog} & \textbf{Stellar/Planet Property} & \textbf{\emph{Overlap}} & \textbf{\emph{UA}} & \textbf{\emph{UB}}\\
\cmidrule{1-5}

\multirow{4}{*}{ EU-OPEN} & $M_{*} $ & 2306 & 185 & 157\\
& $M_{P} $ & \emph{904} & 122 & 60\\

  & $R_{P} $ & \emph{1867} & 105 & 129\\

  & $Period $ & 2268 & 159 & 157\\
\midrule
  \multirow{4}{*}{OPEN-ARCHIVE} & $M_{*} $ & 2290 & 173 & 282\\

 & $M_{P} $ & 877 & 73 & 101\\

  & $R_{P} $ & 1833 & 141 & 218\\

  &$Period $  & 2253 & 171 & 263\\
\midrule
 \multirow{4}{*}{ARCHIVE-EU }& $M_{*} $ & 2348 & 224 & 143\\

 & $M_{P} $ & 1876 & 175 & 72\\

  & $R_{P} $ & 894 & 84 & 113\\

  & $Period $ & 2307 & 209 & 119\\

\bottomrule
\end{tabular}}

    \end{table}
    \unskip
    
    \begin{table}[H]
    \tablesize{\small}
    \centering
    \caption{Stellar mass and planet properties KS test \emph{p-value}s for the different~catalogs.}
    \label{tabref:geosciences-343110-t0A13}

\setlength{\cellWidtha}{\textwidth/5-2\tabcolsep-0in}
\setlength{\cellWidthb}{\textwidth/5-2\tabcolsep+0.4in}
\setlength{\cellWidthc}{\textwidth/5-2\tabcolsep-0.1in}
\setlength{\cellWidthd}{\textwidth/5-2\tabcolsep-0.2in}
\setlength{\cellWidthe}{\textwidth/5-2\tabcolsep-0.2in}
\scalebox{1}[1]{\begin{tabular}{>{\PreserveBackslash\centering}m{\cellWidtha}>{\PreserveBackslash\centering}m{\cellWidthb}>{\PreserveBackslash\centering}m{\cellWidthc}>{\PreserveBackslash\centering}m{\cellWidthd}>{\PreserveBackslash\centering}m{\cellWidthe}}
\toprule

\textbf{Catalog} & \textbf{Stellar/Planet Property} & \textbf{\emph{Overlap} vs. \emph{UA}} & \textbf{\emph{Overlap} vs. \emph{UB}} & \textbf{\emph{UA} vs. \emph{UB}}\\
\cmidrule{1-5}

  \multirow{4}{*}{EU-OPEN}& $M_{*} $ & \textless{}10\textsuperscript{$-$10} & 0.06 & 8 $\times$ 10\textsuperscript{$-$4}\\

 & $M_{P} $ & 6 $\times$ 10\textsuperscript{$-$3} & 0.11 & 0.77\\

  & $R_{P} $ & 2 $\times$ 10\textsuperscript{$-$3} & 9 $\times$ 10\textsuperscript{$-$3} & 1 $\times$ 10\textsuperscript{$-$5}\\

  & $Period $ & 0.02 & 2 $\times$ 10\textsuperscript{$-$8} & 8 $\times$ 10\textsuperscript{$-$6}\\
\midrule
 \multirow{4}{*}{OPEN-ARCHIVE}  & $M_{*} $ & 0.12 & \textless{}10\textsuperscript{$-$10} & 0.01\\

& $M_{P} $ & 0.44 & 0.16 & 0.46\\

  & $R_{P} $ & 0.11 & 0.19 & 0.05\\

  & $Period $ & 0.014 & 0.04 & 0.36\\
\midrule
\multirow{4}{*}{ARCHIVE-EU}  & $M_{*} $ & 2 $\times$ 10\textsuperscript{$-$6} & 5 $\times$ 10\textsuperscript{$-$7} & 0.05\\

 & $M_{P} $ & 0.31 & 0.14 & 0.99\\

  & $R_{P} $ & 0.22 & \textless{}10\textsuperscript{$-$10} & \textless{}10\textsuperscript{$-$10}\\

  & $Period $ & 2 $\times$ 10\textsuperscript{$-$3} & 2 $\times$ 10\textsuperscript{$-$8} & 8 $\times$ 10\textsuperscript{$-$6}\\

\bottomrule
\end{tabular}}

    \end{table}
    \unskip

When comparing the planetary properties of the first detected planets in these systems, we observed a bias towards small planets that were probably easier to detect around lower mass stars. We also found a small fraction of long orbital period planets ($\left. Period \right.\sim 1000~\text{days} $) derived from direct imaging of large, high mass, semi-major axis objects around a small mass star or brown dwarf (especially evident in the EU subsets).

Examining the stellar surface temperature (not displayed here), we found the \emph{unique} vs. \emph{overlap} subsets analysis to be analogous to that of the stellar mass analysis. This was no surprise, especially because of the well-known correlation the two properties possess according to some mass--luminosity relation~\cite{B32-geosciences-343110}. For this reason, we have chosen not to elaborate about it~here.

\unskip

\subsubsection{Stellar Metallicity \label{app:secAdot2dot2-geosciences-343110}}

\begin{figure}[H]
\centering
\includegraphics[scale=1]{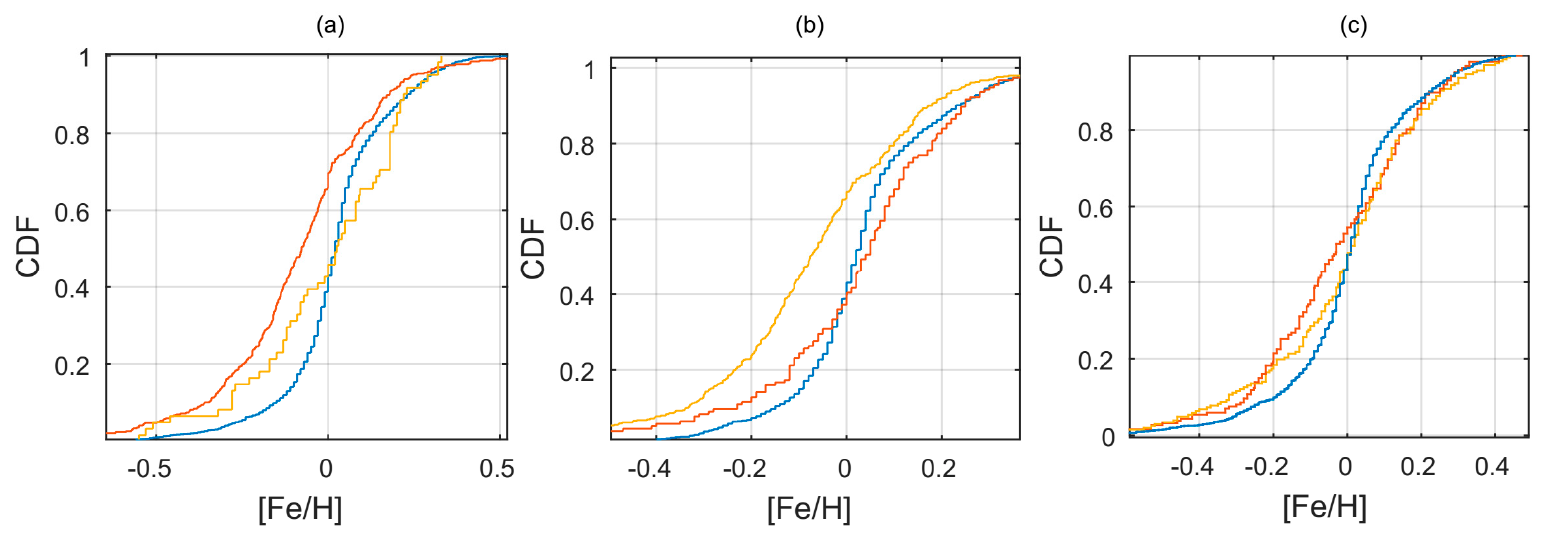}
\caption{CDF of the stellar metallicity property for different subsets. (\textbf{\boldmath{a}}) EU-OPEN; \mbox{(\textbf{\boldmath{b}}) OPEN-ARCHIVE}; (\textbf{\boldmath{c}}) ARCHIVE-EU blue: \emph{Overlap}, red: \emph{UA}, yellow: \emph{UB.}}
\label{fig:geosciences-343110-f0A10}
\end{figure}

\begin{figure}[H]
\centering
\includegraphics[scale=1]{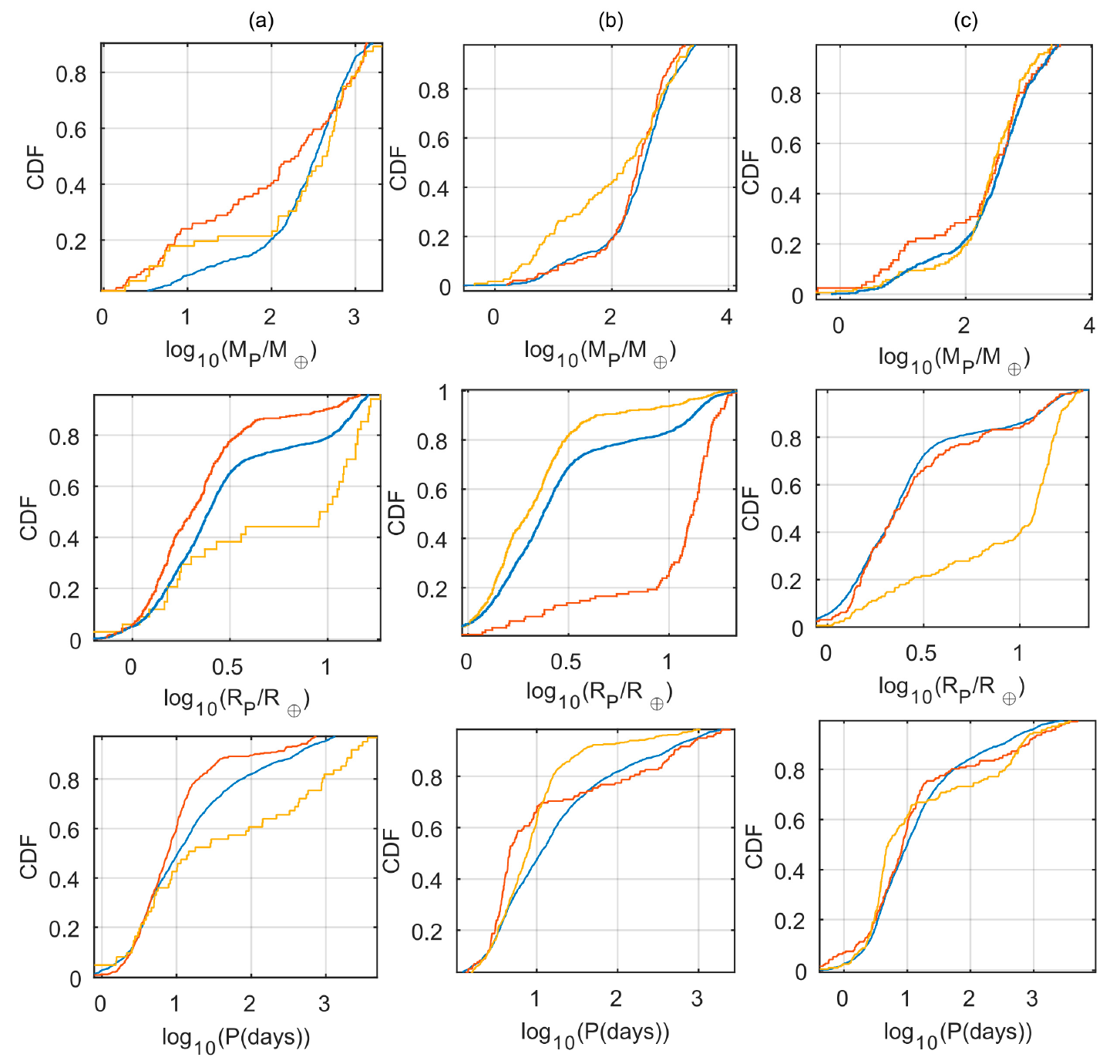}
\caption{CDF of the planetary properties for different subsets of the first detected planet systems with reported stellar metallicity property. (\textbf{\boldmath{a}}) EU-OPEN; (\textbf{\boldmath{b}}) OPEN-ARCHIVE; (\textbf{\boldmath{c}}) ARCHIVE-EU blue: \emph{Overlap}, red: \emph{UA}, yellow: \emph{UB.}}
\label{fig:geosciences-343110-f0A11}
\end{figure}

We found the catalogs usually reported the metallicity property with high relative errors, probably linked to imprecise derivation that is used to determine the metallicity value. Combined with the different survey sources the exoplanet catalogs choose to present in their sites, we consequently found the highest inconsistency between catalogs to be with this parameter (as compared with the other stellar and planetary properties). We present the number of objects with reported stellar metallicity (and first planet’s planetary properties), calculated \emph{p}-values and CDFs of the different subsets in \cref{tabref:geosciences-343110-t0A14,tabref:geosciences-343110-t0A15} and \cref{fig:geosciences-343110-f0A10,fig:geosciences-343110-f0A11}, respectively. We found the \emph{unique} vs. \emph{overlap} subsets to be yet again different. While the \emph{overlap} subsets have a clear peak around solar metallicity, \mbox{the \emph{unique}} subsets have a wider range of metallicities values. Combining information we acquired from the system’s first detected planet, we noted the unique OPEN subset to be populate with a relatively low number of small planets: The OPEN catalog lists fewer Kepler planets with their stellar metallicity, and hence they automatically move to the opposite \emph{unique} subset. We found the \emph{unique} subsets of the ARCHIVE catalog to be especially distributed with a wide variance coming from its population of many K2 and Kepler small radii planets~\cite{B24-geosciences-343110}.

    \begin{table}[H]
    \tablesize{\small}
    \centering
    \caption{Catalogs’ subsets total number of objects with listed stellar metallicity and planetary properties~fractions.}
    \label{tabref:geosciences-343110-t0A14}

\setlength{\cellWidtha}{\textwidth/5-2\tabcolsep-0in}
\setlength{\cellWidthb}{\textwidth/5-2\tabcolsep+0.4in}
\setlength{\cellWidthc}{\textwidth/5-2\tabcolsep-0.6in}
\setlength{\cellWidthd}{\textwidth/5-2\tabcolsep-0.8in}
\setlength{\cellWidthe}{\textwidth/5-2\tabcolsep-0.6in}
\scalebox{0.95}[0.95]{\begin{tabular}{>{\PreserveBackslash\centering}m{\cellWidtha}>{\PreserveBackslash\centering}m{\cellWidthb}>{\PreserveBackslash\centering}m{\cellWidthc}>{\PreserveBackslash\centering}m{\cellWidthd}>{\PreserveBackslash\centering}m{\cellWidthe}}
\toprule

\textbf{Catalog} & \textbf{Stellar/Planet Property} & \textbf{\emph{Overlap}} & \textbf{\emph{UA}} & \textbf{\emph{UB}}\\
\cmidrule{1-5}

  \multirow{4}{*}{EU-OPEN}& $\left\lbrack {Fe/H} \right\rbrack $ & 2035 & 409 & 53\\

 & $M_{P} $ & 798 & 97 & 48\\

  & $R_{P} $ & 1654 & 103 & 24\\

  & $Period $ & 2034 & 407 & 52\\
\midrule
 \multirow{4}{*}{ OPEN-ARCHIVE} & $\left\lbrack {Fe/H} \right\rbrack $ & 1937 & 151 & 487\\

& $M_{P} $ & 691 & 137 & 108\\

  & $R_{P} $ & 1554 & 103 & 450\\

  & $Period $ & 1936 & 150 & 486\\
\midrule
  \multirow{4}{*}{ARCHIVE-EU}& $\left\lbrack {Fe/H} \right\rbrack $ & 2266 & 158 & 178\\

 & $M_{P} $ & 732 & 67 & 144\\

  & $R_{P} $ & 1871 & 133 & 134\\

  & $Period $ & 2264 & 158 & 176\\

\bottomrule
\end{tabular}}

    \end{table}
    \unskip
    
    \begin{table}[H]
    \tablesize{\small}
    \centering
    \caption{Stellar metallicity and planet properties KS test \emph{p-value}s for the different~catalogs.}
    \label{tabref:geosciences-343110-t0A15}

\setlength{\cellWidtha}{\textwidth/5-2\tabcolsep-0in}
\setlength{\cellWidthb}{\textwidth/5-2\tabcolsep+0.4in}
\setlength{\cellWidthc}{\textwidth/5-2\tabcolsep-0.1in}
\setlength{\cellWidthd}{\textwidth/5-2\tabcolsep-0.2in}
\setlength{\cellWidthe}{\textwidth/5-2\tabcolsep-0.2in}
\scalebox{0.95}[0.95]{\begin{tabular}{>{\PreserveBackslash\centering}m{\cellWidtha}>{\PreserveBackslash\centering}m{\cellWidthb}>{\PreserveBackslash\centering}m{\cellWidthc}>{\PreserveBackslash\centering}m{\cellWidthd}>{\PreserveBackslash\centering}m{\cellWidthe}}
\toprule

\textbf{Catalog} & \textbf{Stellar/ Planet Property} & \textbf{\emph{Overlap} vs. \emph{UA}} & \textbf{\emph{Overlap} vs. \emph{UB}} & \textbf{\emph{UA} vs. \emph{UB}}\\
\cmidrule{1-5}

  \multirow{4}{*}{EU-OPEN}& $\left\lbrack {Fe/H} \right\rbrack $ & \textless{}10\textsuperscript{$-$10} & 0.04 & 9 $\times$ 10\textsuperscript{$-$4}\\

 & $M_{P} $ & 5 $\times$ 10\textsuperscript{$-$5} & 0.24 & 0.11\\

  & $R_{P} $ & 2 $\times$ 10\textsuperscript{$-$6} & 8 $\times$ 10\textsuperscript{$-$4} & 3 $\times$ 10\textsuperscript{$-$6}\\

  & $Period $ & \textless{}10\textsuperscript{$-$10} & 2 $\times$ 10\textsuperscript{$-$3} & 10\textsuperscript{$-$5}\\
\midrule
\multirow{4}{*}{OPEN-ARCHIVE } & $\left\lbrack {Fe/H} \right\rbrack $ & 0.02 & \textless{}10\textsuperscript{$-$10} & 3 $\times$ 10\textsuperscript{$-$9}\\

 & $M_{P} $ & 0.13 & 2 $\times$ 10\textsuperscript{$-$5} & 3 $\times$ 10\textsuperscript{$-$4}\\

  & $R_{P} $ & \textless{}10\textsuperscript{$-$10} & 1 $\times$ 10\textsuperscript{$-$6} & \textless{} 10\textsuperscript{$-$10}\\

  & $Period $ & 3 $\times$ 10\textsuperscript{$-$7} & \textless{}10\textsuperscript{$-$10} & 7 $\times$ 10\textsuperscript{$-$5}\\
\midrule
 \multirow{4}{*}{ARCHIVE-EU }& $\left\lbrack {Fe/H} \right\rbrack $ & 2 $\times$ 10\textsuperscript{$-$5} & 0.19 & 10\textsuperscript{$-$3}\\

 & $M_{P} $ & 0.48 & 0.12 & 0.33\\

  & $R_{P} $ & 0.89 & \textless{}10\textsuperscript{$-$10} & \textless{}10\textsuperscript{$-$10}\\

  & $Period $ & 0.05 & 3 $\times$ 10\textsuperscript{$-$6} & 3 $\times$ 10\textsuperscript{$-$3}\\

\bottomrule
\end{tabular}}

    \end{table}
    \unskip

\unskip

\unskip

\subsection{Specific Examples of Differences between the Catalogs \label{app:secAdot3-geosciences-343110}}

Below, we present some of the examples we have come across during this analysis. These examples emphasize some of the differences between the catalogs reported planetary and stellar~properties:
			
\begin{enumerate}
\item[(I)] Planetary mass inconsistency: ‘\emph{Kepler-114 c}’ is a TTV detected mass planet. The EU catalog displays its mass to be $\left. M_{p} \right.\sim 40~M_{\oplus} $~\cite{B26-geosciences-343110}. However, since this measurement is an upper limit, we found it to be inconsistent with the nominal measurement the ARCHIVE displays of $\left. M_{p} \right.\sim 2.8M_{\oplus} $~\cite{B28-geosciences-343110}, while the OPEN catalog does not include this planet; The system of ‘\emph{Kepler 53 b,c}’ is labeled in the ARCHIVE catalog with main planetary mass of $M_{P} $ \textless{} $18~M_{J} $ and \mbox{$M_{P} $ \textless{} $15~M_{J} $}, for planets ‘\emph{b}’, ‘\emph{c}’ respectively~\cite{B25-geosciences-343110}, however a later paper~\cite{B26-geosciences-343110} reports the masses to be much smaller, $\left. M_{P} \right.\sim 0.18~M_{J} $, $\left. M_{P} \right.\sim 0.11~M_{J} $, respectively, as listed in both EU and OPEN. Both EU and OPEN display the mass properties of ‘\emph{Kepler-446 b,c,d}’, however according to the reference article~\cite{B27-geosciences-343110}, the given value is only a coarse estimate for the planets masses and expected semi-amplitudes radial velocities signatures using recent empirically-measured; ‘\emph{Hd181720 b}’ is an object that is listed in the OPEN \emph{unique} with a mass of $\left. M_{p} \right.\sim 0.37~M_{J} $. On the other hand, the EU catalog lists its mass to be $\left. M_{p} \right.\sim 12.12~M_{J} $ (and because of a planet mass larger than our $10~M_{J} $ cutoff, dropped from our analysis). The explanation for this difference is referred in EU to be related with the planetary inclination, measured by astrometry, to be small ($\left. i \right.\sim 1.75\text{$^\circ$} $,~\cite{B33-geosciences-343110}), resulting indeed with a $\left. M_{P}\text{$\cdotp $}sini \right.\sim 0.37~M_{J} $ that was set to be the planet mass in the OPEN~catalog.
\item[(II)] \textls[-20]{Planetary radius inconsistency: ‘\emph{CoRoT-21 b}’ is a $\left. R_{p} \right.\sim 1.3R_{J} $ planet in short orbit (\mbox{$\left. Period \right.\sim 2.7$~days}) which exchanges extreme tidal forces with its parent star~\cite{B34-geosciences-343110}, and is listed in the EU and OPEN catalogs only; On the other hand\emph{,} the radius of ‘\emph{HD 219134 d}’ with a bottom radius limit $> 1.6R_{\oplus} $~\cite{B35-geosciences-343110} is listed only on the ARCHIVE~catalog.}
\item[(III)] Planetary orbital period inconsistency: ‘\emph{51 Eri b}’ is an imaged astrometric giant planet with an assumed orbital period of \emph{period\textasciitilde 14965 days} (\emph{41 years}) listed on both the EU and OPEN catalogs~\cite{B36-geosciences-343110}\textbf{\boldmath{,}} but with missing information in the ARCHIVE, displaying only the semi-major axis measurement~\cite{B37-geosciences-343110}. Another difference between the catalogs for this planet is with its reported planetary mass: $\left. M_{p} \right.\sim 2~M_{J} $ (OPEN and ARCHIVE) $\left. M_{p} \right.\sim 9~M_{J} $ (EU); \emph{’Kepler-37 e}’ is a TTV detected planet reported only on the ARCHIVE and OPEN catalogs with no extra reported information except to each period of \emph{period\textasciitilde 51.19 days}~\cite{B26-geosciences-343110}.
\item[(IV)] Stellar mass inconsistency: ‘\emph{HIP 57050 b}’~\cite{B38-geosciences-343110} is listed in all three catalogs but with a missing stellar mass measurement ($\left. M_{*} \right.\sim 0.34M_{} $) in the ARCHIVE catalog; ‘\emph{OGLE-2014-BLG-1722 b}’ is a $\left. M_{p} \right.\sim 55.3M_{\oplus} $ planet detected by the Microlensing method around a $\left. M_{*} \right.\sim 0.4M_{} $ star (two-planet system), listed only on the EU catalog~\cite{B39-geosciences-343110}.
\end{enumerate}

\reftitle{References}


\begin{thebibliography}{999}
\bibitem{B1-geosciences-343110}
Mayor,~M.; Queloz,~D. A Jupiter-mass companion to a solar-type star. \emph{Nature} \textbf{\boldmath{1995}}, \emph{378}, 355--359. [\href{http://dx.doi.org/10.1038/378355a0}{CrossRef}]

\bibitem{B2-geosciences-343110}
Masset,~F.S.; Papaloizou,~J.C.B. Runaway Migration and the Formation of Hot Jupiters. \emph{Astrophys. J.} \textbf{\boldmath{2003}}, \emph{588}, 494--508. [\href{http://dx.doi.org/10.1086/373892}{CrossRef}]

\bibitem{B3-geosciences-343110}
Mordasini,~C.; Alibert,~Y.; Klahr,~H.; Henning,~T. Characterization of exoplanets from their formation I. Models of combined planet formation and evolution. \emph{Astron. Astrophys.} \textbf{\boldmath{2012}}, \emph{547}, 111. [\href{http://dx.doi.org/10.1051/0004-6361/201118457}{CrossRef}]

\bibitem{B4-geosciences-343110}
Ribas,~I.; Miralda-Escud{\fontencoding{T5}\selectfont{\'e}},~J. The eccentricity-mass distribution of exoplanets: Signatures of different formation mechanisms? \emph{Astron. Astrophys.} \textbf{\boldmath{2007}}, \emph{464}, 779--785. [\href{http://dx.doi.org/10.1051/0004-6361:20065726}{CrossRef}]

\bibitem{B5-geosciences-343110}
Williams,~J.P.; Cieza,~L.A. Protoplanetary Disks and Their Evolution. \emph{Annu. Rev. Astron. Astrophys.} \textbf{\boldmath{2011}}, \emph{49}, 67--117. [\href{http://dx.doi.org/10.1146/annurev-astro-081710-102548}{CrossRef}]

\bibitem{B6-geosciences-343110}
Udry,~S.; Santos,~N.C. Statistical Properties of Exoplanets. \emph{Annu. Rev. Astron. Astrophys.} \textbf{\boldmath{2007}}, \emph{45}, 397--439. [\href{http://dx.doi.org/10.1146/annurev.astro.45.051806.110529}{CrossRef}]

\bibitem{B7-geosciences-343110}
Helled,~R.; Bodenheimer,~P.; Podolak,~M.; Boley,~A.; Meru,~F.; Nayakshin,~S.; Fortney,~J.J.; Mayer,~L.; Alibert,~Y.; Boss,~A.P. Giant Planet Formation, Evolution, and Internal Structure. In \emph{Protostars and Planets VI}; Klessen,~S., Dullemond,~P., Henning,~K., Eds.;  University of Arizona Press: Tucson, AZ, USA, 2014.

\bibitem{B8-geosciences-343110}
Christiansen,~J.L. Exoplanet Catalogues. In \emph{Handbook of Exoplanets}; Deeg,~H.J., Belmonte,~J.A., Eds.;  Springer: New York, NY, USA, 2018; ISBN 978-3-319-30648-3.

\bibitem{B9-geosciences-343110}
Schneider,~J.; Dedieu,~C.; Sidaner,~P.L.; Savalle,~R.; Zolotukhin,~I. Defining and cataloging exoplanets: \mbox{The exoplanet.eu} database. \emph{Astron. Astrophys.} \textbf{\boldmath{2011}}, \emph{532}, A79. [\href{http://dx.doi.org/10.1051/0004-6361/201116713}{CrossRef}]

\bibitem{B10-geosciences-343110}
Akeson,~R.L.; Chen,~X.; Ciardi,~D.; Crane,~M.; Good,~J.; Harbut,~M.; Jackson,~E.; Kane,~S.R.; Laity,~A.C.; Leifer,~S.; et~al. The NASA Exoplanet Archive: Data and Tools for Exoplanet Research. \emph{Publ. Astron. Soc. Pac.} \textbf{\boldmath{2013}}, \emph{125}, 989--999. [\href{http://dx.doi.org/10.1086/672273}{CrossRef}]

\bibitem{B11-geosciences-343110}
Wright,~J.T.; Fakhouri,~O.; Marcy,~G.W.; Han,~E.; Feng,~Y.; Johnson,~J.A.; Howard,~A.W.; Fischer,~D.A.; Valenti,~J.A.; Anderson,~J.; et~al. The Exoplanet Orbit Database. \emph{Publ. Astron. Soc. Pac.} \textbf{\boldmath{2011}}, \emph{123}, 412--422. [\href{http://dx.doi.org/10.1086/659427}{CrossRef}]

\bibitem{B12-geosciences-343110}
Wright,~J.T.; Gaudi,~B.S. Exoplanet Detection Methods. In \emph{Planets, Stars and Stellar Systems}; Springer: Dordrecht, The Netherlands, 2013; pp.~489--540.

\bibitem{B13-geosciences-343110}
Rowe,~J.F.; Bryson,~S.T.; Marcy,~G.W.; Lissauer,~J.J.; Jontof-Hutter,~D.; Mullally,~F.; Gilliland,~R.L.; Issacson,~H.; Ford,~E.; Howell,~S.B.; et~al. Validation of Kepler’s Multiple Planet Candidates. III: Light Curve Analysis and Announcement of Hundreds of New Multi-planet Systems. \emph{Astrophys. J.} \textbf{\boldmath{2014}}, \emph{784}, 45. [\href{http://dx.doi.org/10.1088/0004-637X/784/1/45}{CrossRef}]

\bibitem{B14-geosciences-343110}
Huber,~C.; Silva Aguirre,~V.; Matthews,~J.M.; Pinsonneault,~M.H.; Gaidos,~E.; Garcia,~R.A.; Hekker,~S.; Huber,~D.; Garc{\fontencoding{T5}\selectfont{\'i}}a,~R.A.; Mathur,~S.; et~al. Revised Stellar Properties Of Kepler Targets For The Quarter 1-16 Transit Detection Run Detailed Terms. \emph{Astrophys. J.} \textbf{\boldmath{2014}}, \emph{211}. [\href{http://dx.doi.org/10.1088/0067-0049/211/1/2}{CrossRef}]

\bibitem{B15-geosciences-343110}
Johnson,~J.A.; Petigura,~E.A.; Fulton,~B.J.; Marcy,~G.W.; Howard,~A.W.; Isaacson,~H.; Hebb,~L.; Cargile,~P.A.; Morton,~T.D.; Weiss,~L.M.; et~al. The California-Kepler Survey. II. Precise Physical Properties of 2025 Kepler Planets and Their Host Stars. \emph{Astrophys. J.} \textbf{\boldmath{2017}}, 1--13. [\href{http://dx.doi.org/10.3847/1538-3881/aa80e7}{CrossRef}]

\bibitem{B16-geosciences-343110}
Massey,~F.J. The Kolmogorov-Smirnov Test for Goodness of Fit. \emph{J. Am. Stat. Assoc.} \textbf{\boldmath{1951}}, \emph{46}, 68--78. [\href{http://dx.doi.org/10.1080/01621459.1951.10500769}{CrossRef}]

\bibitem{B17-geosciences-343110}
Silverman,~B.W. Density estimation for statistics and data analysis. In \emph{Monographs on Statistics and Applied Probability}; Chapman and Hall: London, UK, 1986.

\bibitem{B18-geosciences-343110}
Lithwick,~Y.; Xie,~J.; Wu,~Y. Extracting Planet Mass and Eccentricity from TTV data. \emph{Astrophys. J.} \textbf{\boldmath{2012}}, \emph{761}, 122. [\href{http://dx.doi.org/10.1088/0004-637X/761/2/122}{CrossRef}]

\bibitem{B19-geosciences-343110}
Steffen,~J.H. Sensitivity bias in the mass-radius distribution from transit timing variations and radial velocity measurements. \emph{MNRAS} \textbf{\boldmath{2016}}, \emph{457}, 4384--4392. [\href{http://dx.doi.org/10.1093/mnras/stw241}{CrossRef}]

\bibitem{B20-geosciences-343110}
Gao,~P.; Plavchan,~P.; Gagn{\fontencoding{T5}\selectfont{\'e}},~J.; Furlan,~E.; Bottom,~M.; Anglada-Escud{\fontencoding{T5}\selectfont{\'e}},~G.; White,~R.; Davison,~C.; Beichman,~C.; Brinkworth,~C.; et~al. Retrieval of Precise Radial Velocities from Near-Infrared High Resolution Spectra of Low Mass Stars. \emph{Publ. Astron. Soc. Pac.} \textbf{\boldmath{2016}}, \emph{128}, 104501. [\href{http://dx.doi.org/10.1088/1538-3873/128/968/104501}{CrossRef}]

\bibitem{B21-geosciences-343110}
Boss,~A.P. Stellar Metallicity and the Formation of Extrasolar Gas Giant Planets. \emph{Astrophys. J.} \textbf{\boldmath{2002}}, \emph{567}, L149--L153. [\href{http://dx.doi.org/10.1086/340108}{CrossRef}]

\bibitem{B22-geosciences-343110}
Fischer,~D.A.; Valenti,~J. The Planet-Metallicity Correlation. \emph{Astrophys. J.} \textbf{\boldmath{2005}}, \emph{622}, 1102--1117. [\href{http://dx.doi.org/10.1086/428383}{CrossRef}]

\bibitem{B23-geosciences-343110}
Sousa S{\fontencoding{T5}\selectfont{\'e}}rgio,~G.; Santos Nuno,~C.; Israelian,~G.; Mayor,~M.; Stephane,~U. Spectroscopic stellar parameters for 582 FGK stars in the HARPS volume-limited sample-Revising the metallicity-planet correlation. \emph{\mbox{Astron. Astrophys.}} \textbf{\boldmath{2011}}, \emph{533}, A141. [\href{http://dx.doi.org/10.1051/0004-6361/201117699}{CrossRef}]

\bibitem{B24-geosciences-343110}
Buchhave,~L.A.; Latham,~D.W.; Johansen,~A.; Bizzarro,~M.; Torres,~G.; Rowe,~J.F.; Batalha,~N.M.; Borucki,~W.J.; Brugamyer,~E.; Caldwell,~C.; et~al. An abundance of small exoplanets around stars with a wide range of metallicities. \emph{Nature} \textbf{\boldmath{2012}}, \emph{486}, 375--377. [\href{http://dx.doi.org/10.1038/nature11121}{CrossRef}] [\href{http://www.ncbi.nlm.nih.gov/pubmed/22722196}{PubMed}]

\bibitem{B25-geosciences-343110}
Ford,~E.B.; Fabrycky,~D.C.; Steffen,~J.H.; Carter,~J.A.; Fressin,~F.; Holman,~M.J.; Lissauer,~J.J.; Moorhead,~A.V.; Morehead,~R.C.; Ragozzine,~D.; et~al. Transit Timing Observations from Kepler: II. Confirmation of two multiplanet systems via a non-parametric correlation. \emph{Astrophys. J.} \textbf{\boldmath{2012}}, \emph{750}, 113.

\bibitem{B26-geosciences-343110}
Hadden,~S.; Lithwick,~Y. Densities and eccentricities of 139 Kepler planets from transit time variations. \emph{Astrophys. J.} \textbf{\boldmath{2014}}, \emph{787}, 80. [\href{http://dx.doi.org/10.1088/0004-637X/787/1/80}{CrossRef}]

\bibitem{B27-geosciences-343110}
Muirhead,~P.S.; Mann,~A.W.; Vanderburg,~A.; Morton,~T.D.; Kraus,~A.; Ireland,~M.; Swift,~J.J.; Feiden,~G.A.; Gaidos,~E.; Gazak,~J.Z. Kepler-445, Kepler-446 and the Occurrence of Compact Multiples Orbiting Mid-M Dwarf Stars. In \emph{AAS/Division for Extreme Solar Systems Abstracts}; American Astronomical Society: Washington, DC, USA, 2015; Volume 3.

\bibitem{B28-geosciences-343110}
Xie,~J.W. Transit timing variation of near-resonance planetary pairs. II. Confirmation of 30 planets in \mbox{15 multiple} planet systems. \emph{Astrophys. J. Suppl. Ser.} \textbf{\boldmath{2013}}, \emph{210}, 25. [\href{http://dx.doi.org/10.1088/0067-0049/210/2/25}{CrossRef}]

\bibitem{B29-geosciences-343110}
Batalha,~N.M. Exploring exoplanet populations with NASA’s Kepler Mission. \emph{Proc. Nati. Acad. Sci. USA} \textbf{\boldmath{2014}}, \emph{111}, 12647--12654. [\href{http://dx.doi.org/10.1073/pnas.1304196111}{CrossRef}] [\href{http://www.ncbi.nlm.nih.gov/pubmed/25049406}{PubMed}]

\bibitem{B30-geosciences-343110}
Bashi,~D.; Helled,~R.; Zucker,~S.; Mordasini,~C. Two empirical regimes of the planetary mass-radius relation. \emph{Astron. Astrophys.} \textbf{\boldmath{2017}}, \emph{604}, A83. [\href{http://dx.doi.org/10.1051/0004-6361/201629922}{CrossRef}]

\bibitem{B31-geosciences-343110}
Lovis,~C.; Fischer,~D.A. Radial Velocity. In \emph{Radial Velocity Techniques for Exoplanets}; Seager,~S., Ed.;  University of Arizona Press: Tucson, AZ, USA, 2011.

\bibitem{B32-geosciences-343110}
Kuiper,~G.P. The Empirical Mass-Luminosity Relation. \emph{Astrophys. J.} \textbf{\boldmath{1938}}, \emph{88}, 472. [\href{http://dx.doi.org/10.1086/143999}{CrossRef}]

\bibitem{B33-geosciences-343110}
Reffert,~S.; Quirrenbach,~A. Mass constraints on substellar companion candidates from the re-reduced Hipparcos intermediate astrometric data: Nine confirmed planets and two confirmed brown dwarfs. \emph{\mbox{Astron. Astrophys.}} \textbf{\boldmath{2011}}, \emph{527}, A140. [\href{http://dx.doi.org/10.1051/0004-6361/201015861}{CrossRef}]

\bibitem{B34-geosciences-343110}
Pätzold,~M.; Endl,~M.; Csizmadia,~S.; Gandolfi,~D.; Jorda,~L.; Grziwa,~S.; Carone,~L.; Pasternacki,~T.; Aigrain,~S.; Almenara,~J.M.; et~al. Transiting exoplanets from the CoRoT space mission XXIII. CoRoT-21b: A doomed large Jupiter around a faint subgiant star. \emph{Astron. Astrophys.} \textbf{\boldmath{2012}}, \emph{545}, 6. [\href{http://dx.doi.org/10.1051/0004-6361/201118425}{CrossRef}]

\bibitem{B35-geosciences-343110}
Gillon,~M.; Demory,~B.-O.; Van Grootel,~V.; Motalebi,~F.; Lovis,~C.; Cameron,~A.C.; Charbonneau,~D.; Latham,~D.; Molinari,~E.; Pepe,~F.A.; et~al. Two massive rocky planets transiting a K-dwarf 6.5 parsecs away. \emph{Nat. Astron.} \textbf{\boldmath{2017}}, 1. [\href{http://dx.doi.org/10.1038/s41550-017-0056}{CrossRef}]

\bibitem{B36-geosciences-343110}
De Rosa,~R.J.; Nielsen,~E.L.; Blunt,~S.C.; Graham,~J.R.; Konopacky,~Q.M.; Marois,~C.; Pueyo,~L.; Rameau,~J.; Ryan,~D.M.; Wang,~J.J.; et~al. Astrometric Confirmation and Preliminary Orbital Parameters of the Young Exoplanet 51 Eridani b with the Gemini Planet Imager. \emph{Astrophys. J.} \textbf{\boldmath{2015}}, 814. [\href{http://dx.doi.org/10.1088/2041-8205/814/1/l3}{CrossRef}]

\bibitem{B37-geosciences-343110}
Macintosh,~B.; Graham,~J.R.; Barman,~T.; De Rosa,~R.J.; Konopacky,~Q.; Marley,~M.S.; Marois,~C.; Nielsen,~E.L.; Pueyo,~L.; Rajan,~A.; et~al. Discovery and spectroscopy of the young Jovian planet 51 Eri b with the Gemini Planet Imager. \emph{Sci. Express} \textbf{\boldmath{2015}}, \emph{350}, 64--67. [\href{http://dx.doi.org/10.1126/science.aac5891}{CrossRef}] [\href{http://www.ncbi.nlm.nih.gov/pubmed/26272904}{PubMed}]

\bibitem{B38-geosciences-343110}
Haghighipour,~N.; Vogt,~S.S.; Paul Butler,~R.; Rivera,~E.J.; Laughlin,~G.; Meschiari,~S.; Henry,~G.W. \mbox{The Lick-Carnegie} Exoplanet Survey: A Saturn-Mass Planet in the Habitable Zone of the Nearby M4V Star HIP 57050. \emph{Astrophys. J.} \textbf{\boldmath{2010}}, \emph{715}, 271--276. [\href{http://dx.doi.org/10.1088/0004-637X/715/1/271}{CrossRef}]

\bibitem{B39-geosciences-343110}
Suzuki,~D.; Bennett,~D.P.; Udalski,~A.; Bond,~I.A.; Sumi,~T.; Han,~C.; Abe,~F.; Asakura,~Y.; Barry,~R.K.; Bhattacharya,~A.; et~al. A Likely Detection of a Two-Planet System in a Low Magnification Microlensing Event. \emph{Astrophys. J.} \textbf{\boldmath{2018}}, \emph{155}, 14. [\href{http://dx.doi.org/10.3847/1538-3881/aabd7a}{CrossRef}]

\end{thebibliography}
\end{document}